\documentclass[5p, preprint]{elsarticle}
\usepackage{boxedminipage}
\usepackage{nopageno}
\usepackage{graphicx}
\usepackage{natbib}
\usepackage[utf8]{inputenc}
\usepackage[T1]{fontenc}
\usepackage{textcomp}
\usepackage{siunitx}
\usepackage[usenames,dvipsnames]{xcolor}
\usepackage{listings, lineno}
\usepackage{caption, colortbl}
\usepackage{subcaption, amsthm, amsmath, amsfonts, amssymb}
\usepackage{multirow}
\usepackage{algorithm}
\usepackage{placeins}
\usepackage{algpseudocode, balance}
\definecolor{LightCyan}{rgb}{0.88,1,1}
\definecolor{BadGrey}{rgb}{0.98,0.88,0.99}

\definecolor{color_19}{HTML}{AFEEEE}
\definecolor{color_20}{HTML}{1F77B4}
\definecolor{color_21}{HTML}{FF7F0E}
\definecolor{color_22}{HTML}{D62728}

\usepackage{graphicx}
\usepackage{amssymb}
\usepackage{amsthm}

\biboptions{square,sort&compress,numbers}

\journal{}

\usepackage{framed} 

\usepackage{multicol} 

\usepackage{nomencl} 

\usepackage{lipsum}

\makenomenclature

\renewcommand*\nompreamble{\begin{multicols}{2}}

\renewcommand*\nompostamble{\end{multicols}}

\begin{document}

\begin{frontmatter}


\title{Experimental Data-Driven Model Predictive Control \\ of a Hospital HVAC System During Regular Use}


\author[EPFL]{Emilio T. Maddalena}
\author[EPFL]{Silvio A. Müller}
\author[HSJ]{Rafael M. dos Santos}
\author[EPFL]{Christophe Salzmann}
\author[EPFL,corr]{Colin N. Jones}

\cortext[corr]{Corresponding author e-mail: \texttt{colin.jones@epfl.com}.}

\address[EPFL]{Laboratoire d’Automatique, \'{E}cole Polytechnique F\'{e}d\'{e}rale de Lausanne (EPFL), Lausanne, Switzerland}
\address[HSJ]{Hospital São Julião, Campo Grande, Brazil}

\begin{abstract}
    Herein we report a multi-zone, heating, ventilation and air-conditioning (HVAC) control case study of an industrial plant responsible for cooling a hospital surgery center. The adopted approach to guaranteeing thermal comfort and reducing electrical energy consumption is based on a statistical non-parametric, non-linear regression technique named Gaussian processes. Our study aimed at assessing the suitability of the aforementioned technique to learning the building dynamics and yielding models for our model predictive control (MPC) scheme. Experimental results gathered while the building was under regular use showcase the final controller performance while subject to a number of measured and unmeasured disturbances. Finally, we provide readers with practical details and recommendations on how to manage the computational complexity of the on-line optimization problem and obtain high-quality solutions from solvers.
\end{abstract}


\end{frontmatter}


    
    
    
    
\section{Introduction}

Although it is rather common to find studies claiming that advanced building control techniques paired with modern machine learning models can enhance the system operation and attain important monetary savings, it is considerably harder to find significant experimental evidence supporting these statements \cite{maddalena2020data,drgovna2020all,hong2020state,zhang2021review}. Instead, the validation of complex analysis and control techniques is usually performed with the aid of simulation models calibrated either on real data or on construction parameters \cite{domahidi2014learning,terzi2020learning,chakrabarty2021accelerating,huang2021simulation,buttitta2021evaluation}. Some investigations adopt a middle ground approach, whereby novel strategies are tested through extensive simulations followed by scaled-down experiments acting as a proof of concept \cite{aswani2011reducing,ma2014stochastic,svetozarevic2021data}. Simulations are at times the only way of exciting buildings with aggressive signals or assessing their behavior under very specific and reproducible weather conditions. Nevertheless, in our view the field can still greatly benefit from more experimental validations of certain machine learning tools present in the literature that hold significant promise  \cite{drgovna2018approximate,ding2019octopus,li2021model}. 

\textbf{Related work:} A detailed investigation of the benefits of using Regression Trees and Random Forests as modeling techniques for buildings is presented in \cite{smarra2018data}, where the authors pair them with MPC and show their effectiveness in a number of interesting case studies, including a demand-response setting. Similarly, Random Forests are investigated in \cite{bunning2020experimental} as an easy-to-use and maintain modeling strategy for predictive control of buildings, aiming at reducing their energy footprint. The residential apartment considered in the study is a part of the NEST testbed located in Switzerland and features water-based ceiling panels for both heating and cooling. The same building platform was used in a distributed MPC investigation recently reported in \cite{lefebure2021distributed}, which targets the large-scale problem of coordinating energy hubs. The experimental section of the work however only reports real experiments on a single zone over a total period of 24 hours. Feedforward neural networks (NN) are another popular machine learning tool in the building and HVAC fields \cite{huang2015neural,afram2017artificial,yang2020model,di2021physically,yang2022machine}. Given an appropriate architecture, these models certainly have great representation capabilities and are nowadays rather straightforward to train thanks to the availability of reliable frameworks such as \texttt{Pytorch} \cite{paszke2019pytorch} and \texttt{TensorFlow} \cite{abadi2016tensorflow}. The known downside of NNs is their low sample efficiency, in that the batches of data required to learn a given physical phenomenon is large when compared to non-parametric alternatives.

\begin{figure*}[!t]
    \centering
    \includegraphics[width=0.492\linewidth]{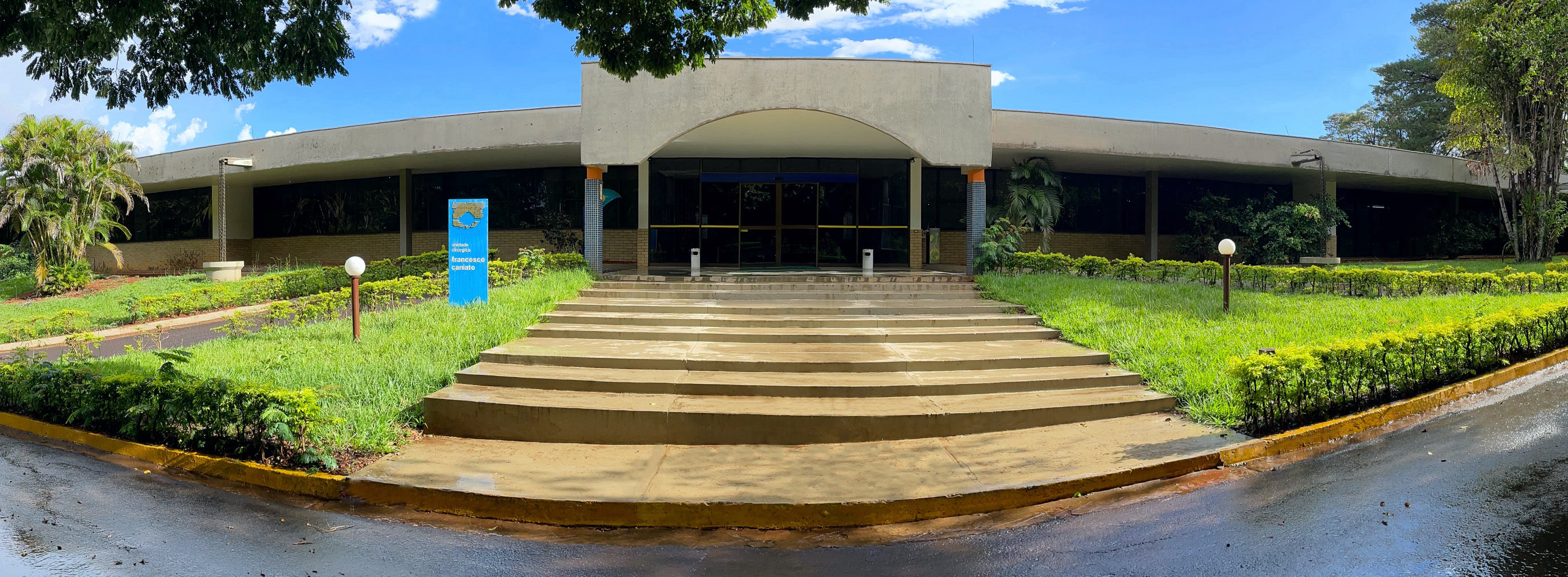} \hspace{0.25pt}
    \includegraphics[width=0.492\linewidth]{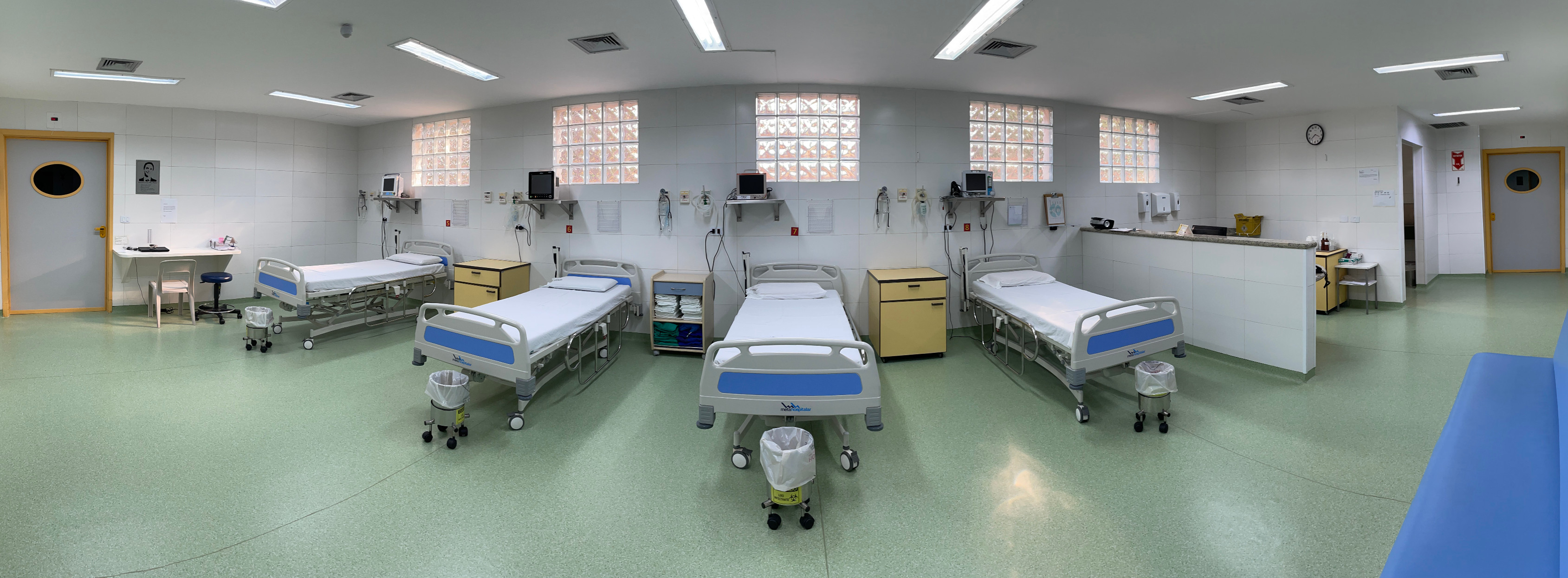}
    \caption{The facade of the Hospital São Julião surgery center, situated in Campo Grande, MS, Brazil (left). The three thermal zones considered in this study (right): The waiting room and the doors that lead to the two operating rooms, all in the ophthalmology section of the building, located on its West end.}
    \label{fig.facadeAndRooms}
\end{figure*}

Gaussian processes (GPs) are a non-linear modeling strategy grounded in Bayesian statistics \cite{williams2006gaussian}. As opposed to the aforementioned approaches that compress information into a predetermined number of parameters (e.g. the weights of a NN whose architecture has been chosen), the expressive power of GPs grows indefinitely as more data is used to train them. Besides their great flexibility, they are also appealing for yielding confidence intervals that quantify the uncertainty associated with their predictions. This machine learning technique has been used in the building domain to calibrate complex simulations \cite{chong2017bayesian,chakrabarty2021}, replace grey-box RC models \cite{gray2016thermal}, forecast energy consumption \cite{yoon2018energy,zeng2020prediction}, detect faults in HVAC equipment \cite{van2017advanced} and create abstract models for how consumers react to demand-adjustment signals \cite{nghiem2017data}. When combined with optimization strategies however, such non-parametric models typically incur a considerable computational load. More specifically, one has to deal with non-convex programs with potentially many local optima in real-time. Since building automation systems have fairly long sampling periods and are usually deployed on general purpose computers, they could in principle support a Gaussian process-based MPC scheme. Yet, to the best of our knowledge, no experimental investigation of this approach considering multiple zones has been reported in the literature so far.

\textbf{Contributions:} In this paper we assess the suitability of using risk-aware non-parametric models combined with MPC to tackle an HVAC control problem. More specifically, the industrial cooling system of a hospital located in Brazil was considered, and a Gaussian process-based MPC controller was designed and deployed to guarantee thermal comfort to the occupants while minimizing the associated electrical demand\footnote{The complete project dataset and its documentation are publicly available at \texttt{https://gitlab.epfl.ch/la/la-database}.}. The following factors being simultaneously present in the current study distinguish it largely from previous ones:
\begin{itemize}
    \item The investigation was concerned with three thermal zones, all having external walls and one being affected by considerable direct solar radiation.
    \item The rooms were fully occupied during working hours (from 7 am to 7 pm), implying a large inflow and outflow of people and internal heat gain variations. 
    \item The MPC controller was not restricted to generating set-points for low-level devices, but had full control over the air-handling unit valves.
    \item The building dynamics were learned by Gaussian processes, whose uncertainty estimates were used to robustify the MPC controller.
\end{itemize}
We also detail practical aspects of the project in terms of hardware and thoroughly discuss the main challenges faced. The computational complexity of the non-convex real-time optimization algorithm is analyzed along with its scalability potential. 

The paper is organized as follows. Section~\ref{sec.Preliminaries} is devoted to describing the physical aspects of the building, its HVAC system and the local hardware used in the project. Next, in Section~\ref{sec.Methodology}, we analyze the control problem and detail the employed modelling techniques, which culminates in our MPC formulation given in \eqref{eq.optControlProb}. The reader with enough control background might benefit from directly studying \eqref{eq.optControlProb} rather than reading Section~\ref{sec.Methodology} sequentially. Finally, simulation and experimental results are shown in Section~\ref{sec.SimExpRes}, which are then followed by our conclusions.

\section{Preliminaries}
\label{sec.Preliminaries}

\subsection{The building and its HVAC system}

The building considered in this study was a surgery center situated in the São Julião hospital complex, in the city of Campo Grande, MS, Brazil (Figure~\ref{fig.facadeAndRooms}, left). The 51 rooms that compose it are in permanent use and, for information purposes, 528 surgical procedures were carried out in it during August 2021. We were concerned with three thermal zones in its ophthalmology section: two operating rooms (ORs) and one waiting room (WR), all located on the West end of the building (Figure~\ref{fig.facadeAndRooms}, right). Whereas the former rooms are only connected to the waiting room, the latter has a door to the rest of the surgery center. Opaque glass bricks are present in the waiting room as can be seen in the picture, allowing some natural light to enter the space; the operating rooms on the other hand do not feature them, nor do they have any windows. All spaces have exterior walls, but the right-hand side operating room is significantly more affected by direct solar radiation due to the disposition of the nearby trees.

A forced-air HVAC plant is in place to provide the occupants with a suitable indoor climate in accordance with local regulations. A total of seven air-handling units (AHUs) collect outdoor air that is then treated and filtered before being pumped into the several indoor spaces. We had control only over three AHUs, one for each aforementioned thermal zone. A central chiller connected to an external cooling tower provides chilled water to all AHUs, which in turn feature three-way valves to control the flow of water through their cooling coils. The AHU fans are operated always at constant speed, resulting in a constant volumetric flow through the air-ducts and into the zones. As per the regulations, no air recycling is possible and all return air is directly discharged into the atmosphere. As the temperature in Campo Grande is typically high, the HVAC system was conceived to only cool the space, not having the means to provide positive thermal energy (for more details, see Section~\ref{sec.controlProb}).

\subsection{Sensor networks and the local hardware}

\begin{figure}[!t]
    \centering
    \includegraphics[width=0.9\linewidth]{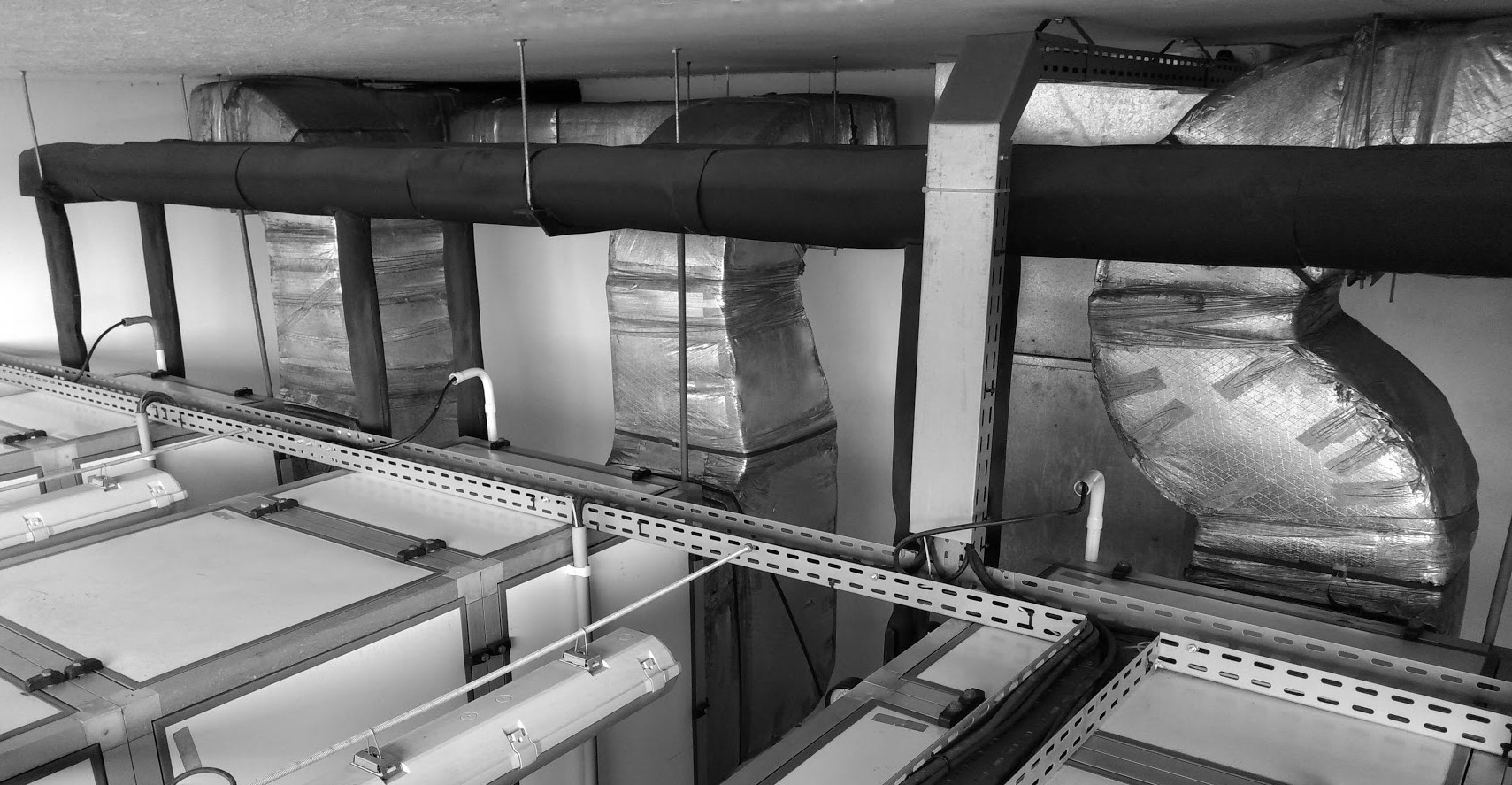} \\[4pt]
    \includegraphics[width=0.9\linewidth]{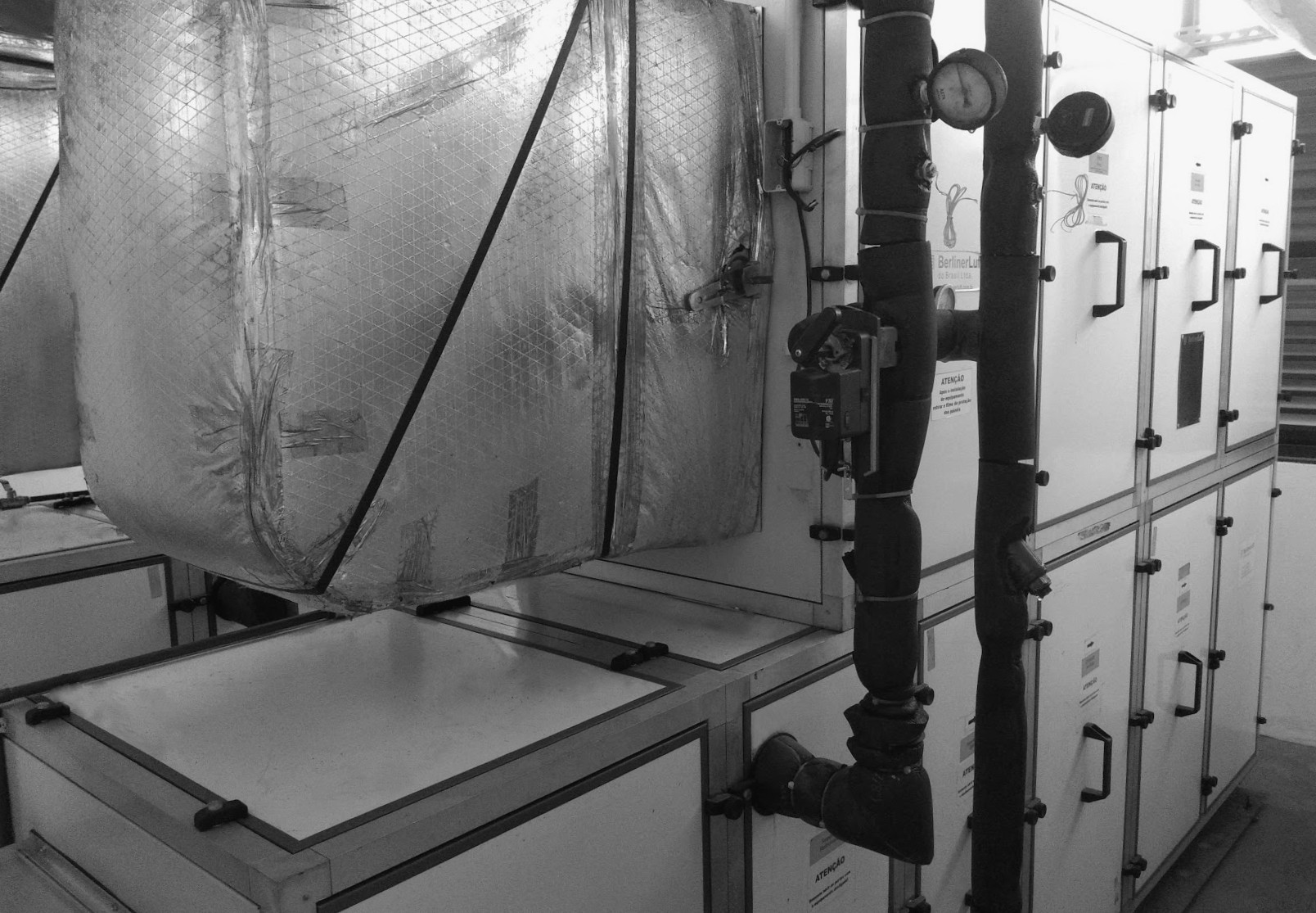} 
    \caption{Photos of the AHU room depicting the air ducts (top), the supply and return water pipes (top and bottom), and the three-way valve servomotor (bottom).}
    \label{fig.ahuRoom}
    \vspace{4pt}
    \includegraphics[width=0.98\linewidth]{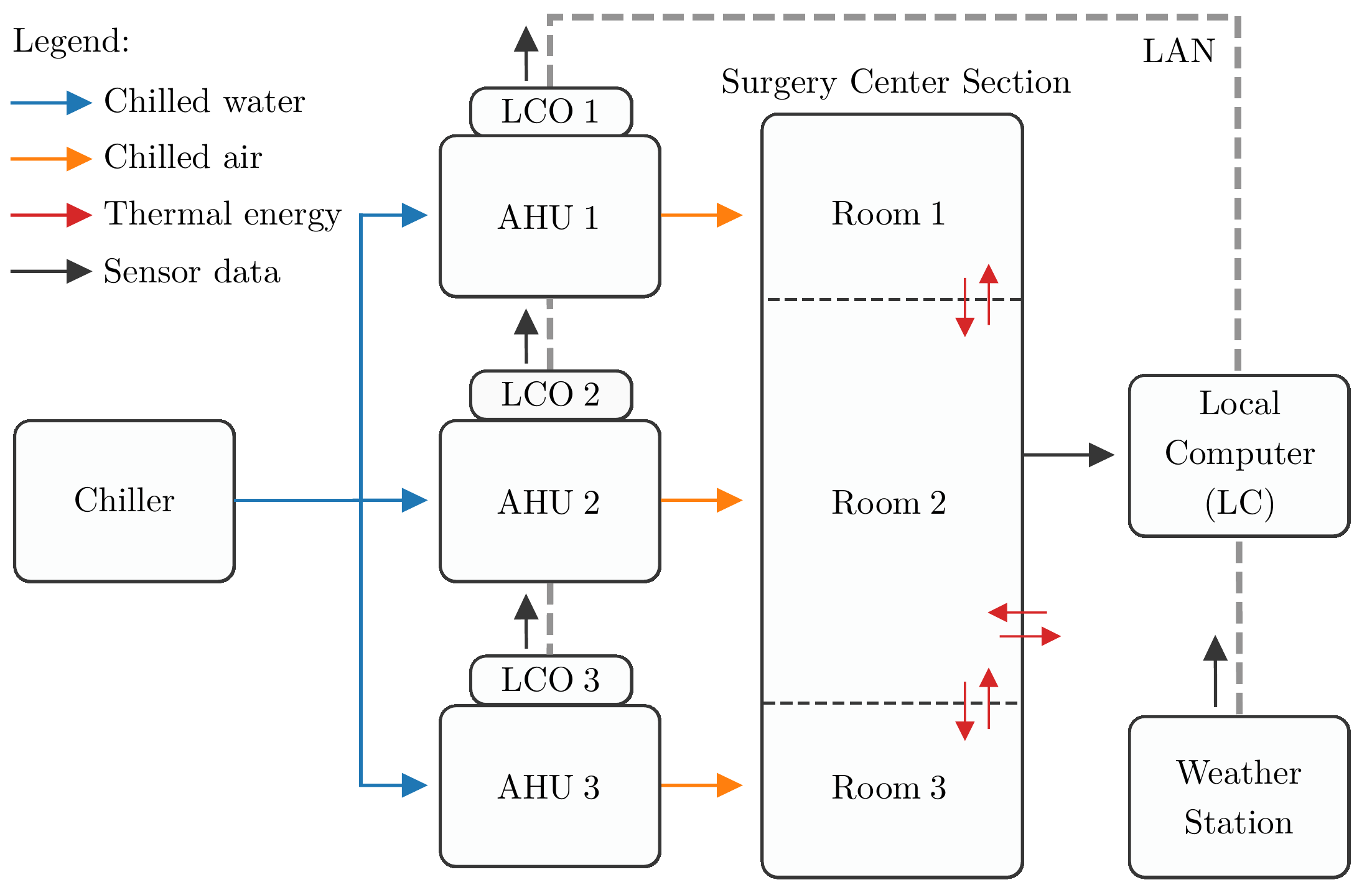} 
    \caption{The overall system architecture, including the three air-handling units (AHUs) and their local controllers (LCOs). Information is exchanged between the LCOs and the local computer (LC) through a local area network (LAN).}
    \label{fig.blockDiagram}
\end{figure}

Two distinct sensor networks were deployed to monitor the HVAC plant and the indoor spaces. Firstly, we will describe the one located in the AHU room. One local controller (LCO)--a National Instruments myRIO--was attached to each air-handling unit, reading all sensors used to monitor the AHUs: supply and return water temperature probes, a water flow meter, an anemometer, as well as an angular position sensor. The LCOs were moreover responsible for running low-level signal processing routines and implementing control actions, i.e., acting on the three-way valve servomotor to change the chilled water flow, hence influencing the supply air temperature. Photos of the AHU room are shown in Figure~\ref{fig.ahuRoom}. Next, in order to measure the indoor temperatures in a flexible way, a wireless network of Z-wave sensors was set up in the operating rooms and waiting room. These were equipped with external temperature probes (Dallas DS18B20) to guarantee fast and precise readings, reporting their measurements periodically to a local computer (LC) that featured a Z-wave transceiver attached to it.

A 3~GHz, 16 GB RAM, core i7 machine was installed in the waiting room, acting as the main computer platform for the project, i.e., the LC. This computer and the AHU LCOs were all connected to a local area network to exchange information, which was done by using the UDP protocol at a rate of approximately 1~Hz. Lastly, a weather station was deployed on site to measure the outdoor temperature and the solar radiation acting on the building with high accuracy. All signals were sampled with a period of two mins and stored into a local time-series database, InfluxDB. A block-diagram of the complete system is depicted in Figure~\ref{fig.blockDiagram}.  

\section{Methodology}
\label{sec.Methodology}

\subsection{Analysis of the control problem}
\label{sec.controlProb}

The control goal is to regulate the indoor temperature within the three zones ($T_i$, $i=1,2,3$), keeping it always below a pre-specified value $T_{\text{max}}$. Although defining two-level temperature envelopes is common for residences and offices (see e.g. \cite{fiorentini2017hybrid,lian2021adaptive}), some employees still make use of the surgery center spaces during nighttime and, thus, the indoor temperature has to stay below $T_{\text{max}}$ even then. Furthermore, this is to be done while minimizing the chiller energy consumption as dictated by its coefficient of performance (COP) curve and the building thermal load. The controlled variables are the angular positions of each AHU three-way valve ($\theta_i$, $i=1,2,3$) that regulate the flow of water across their cooling coils. Naturally, these quantities are physically limited between a minimum $\theta_\text{min}$ and a maximum value $\theta_\text{max}$.

Several disturbances both of internal and external nature act on the system. The measured ones include the outdoor temperature $T_\text{out}$, the solar radiation $R_\text{sol}$, and the temperature of the water supplied by the chiller to the AHUs $T_\text{sup}$. The variables $T_\text{out}$ and $R_\text{sol}$ directly affect the indoor climate by heating the external walls. $T_\text{out}$ and $T_\text{sup}$ can be regarded as an input disturbance; indeed, these quantities define the HVAC system actuation capabilities along with the valve positions $\theta_i$. The unmeasured disturbances are the internal heat gains generated by occupants and equipment, as well as the eventual opening and closing of doors that lead to air mix among rooms. A summary of the relevant control information described here can be found in Table~\ref{tab.controlInformation}.

\begin{table*}[!ht]
    \small
    \centering
    \begin{tabular}{l|c|c}
    \hline & Symbols & Description \\\hline
    Inputs & $\theta_\text{1}$, $\theta_\text{2}$, $\theta_\text{3}$ & Valve position of AHU 1, AHU 2 and AHU 3 \\
    Outputs & $T_\text{1}$, $T_\text{2}$, $T_\text{3}$ & Temperatures within zone 1, zone 2 and zone 3 \\
    Measured disturbances & $T_\text{sup}$, $T_\text{out}$, $R_\text{sol}$ & Chiller supply water temperature, outdoor temperature, solar radiation \\
    Unmeasured disturbances & $-$ & Internal heat gains (e.g. occupants), opening and closing of the doors \\
    \hline
    \end{tabular}
    \caption{The main physical quantities influencing the HVAC plant and the temperature dynamics inside the rooms.}
    \label{tab.controlInformation}
\end{table*}

\begin{table}[!t]
    \small
    \centering
    \begin{tabular}{l|c|c|c|c|c|c|c|c}
    \hline & $T_\text{1}$ & $T_\text{2}$ & $T_\text{3}$ & $\theta_\text{1}$ & $\theta_\text{2}$ & $\theta_\text{3}$ & $T_\text{sup}$ & $T_\text{out}$ \\\hline
    GP1 (OR 1) & 2 & 1 & $-$ & 1 & $-$ & $-$ & 1 & 1 \\
    GP2 (WR) & 1 & 2 & 1 & $-$ & 1 & $-$ & 1 & 1 \\
    GP3 (OR 2) & $-$ & 1 & 2 & $-$ & $-$ & 1 & 1 & 1 \\
    \hline
    \end{tabular}
    \caption{The signals that composed the feature vector of each Gaussian process, and their respective delay parameters $l$. The symbol ``$-$'' indicates what signals were neglected.}
    \label{tab.features}
\end{table}

Creating a reliable model for the system dynamics is crucial to attain high-performance with Model Predictive Control. In the current setting, this task is not trivial as certain disturbances and control variables enter the dynamics non-linearly. For instance, the globe water valves in the AHUs are not linear actuators in that the flow is not directly proportional to the angular position. For this reason, we decided to adopt a flexible class of statistical models to tackle the modeling problem while requiring as little expert knowledge as possible. Their main advantage being that this class not only predicts the expected system behavior, but also quantifies the uncertainty associated with its predictions, hence allowing for a more robust, risk-aware operation.

\subsection{Crafting Gaussian process dynamical models}

Gaussian processes lie in the class of non-parametric\footnote{Despite the name, non-parametric models still have internal \textit{hyperparameters} to be tuned. Although subtle, this difference has significant practical implications \cite{williams2006gaussian}.}, non-linear, Bayesian models. For a thorough presentation of the topic, we refer the reader to \cite{williams2006gaussian,duvenaud2014automatic}. Their main appeal over other types of statistical modeling paradigms is the analytical tractability, whereby expressions used during training and prediction have closed forms, dispensing with the need of using sample-based approximation techniques (see for example \cite{chong2017comparison,chong2019continuous}). 
Assume one wants to model a given phenomenon $f(x)$ through noisy observations of the form $y = f(x) + \varepsilon$, where $\varepsilon \sim \mathcal{N}(0,\sigma^2_\varepsilon)$ is a zero-mean Gaussian noise of unknown variance. As per the usual Bayesian approach, we define a \textit{prior} model $f(x) \sim \mathcal{N}(m(x),k(x,x))$ and, after gathering some experimental data $D= \{x_n,y_n\}_{n=1}^N$, it is possible to update our beliefs and form a \textit{posterior} model whose point-wise mean and variance are respectively
\begin{subequations}
\begin{align}
    \mu(x) &= m(x) + k_X(x)^\top (K+\sigma^2_{\varepsilon}I)^{-1}(y-m_X) \label{eq.GPmean}\\
    \text{var}(x) &= k(x,x) - k_X(x)^\top (K+\sigma^2_{\varepsilon}I)^{-1} k_X(x) \label{eq.GPvar}
\end{align}
\end{subequations}
where $X$ and $y$ denote the collection of all data features and labels in the dataset $D$, and $k(x,x')$ is the kernel function. $k_{X}(x)$ and $K$ are respectively a column vector and a square matrix of kernel evaluations at $X$ and $x$. Lastly, $I$ represents the identity matrix.

Fully specifying a GP regression model amounts to i) picking a suitable mean function and a suitable non-linearity, i.e., a kernel function $k(x,x)$; and ii) optimizing all model hyperparameters. In our case study, a linear mean $m(x) = Ax + b$ was employed. Among the many kernel maps available in the literature \cite{duvenaud2014automatic}, we chose the anisotropic squared-exponential function, a very popular alternative due to its smoothness and expressive power \cite{lederer2021impact,van2021learning}. This kernel has the form 
\begin{equation}
    k_{\text{SE}}(x,x') = \sigma^2 \exp \left(-\frac{1}{2} \sum_{i=1}^d \left(\frac{x_i - x'_i}{\ell_i}\right)^2\right)
    \label{eq.SEARD}
\end{equation}
where $x_i$ is the $i$th component of the feature vector $x$. In \eqref{eq.SEARD}, $\sigma$ is the so called vertical scale hyperparameter and $\ell_i$ are the horizontal scale (a.k.a. lengthscale) hyperparameters. As for optimizing the constants $A$, $b$, $\sigma$, $\sigma_\varepsilon$ and $\ell_i$, we made use of the log-marginal likelihood objective \cite[Chapter 5]{williams2006gaussian} and a gradient-based procedure. This is a widely adopted criterion, known to contain a regularization term that combats overfitting.

In order to design dynamic models for the room temperatures, we used an auto-regressive approach, meaning that future predictions of a signal depend on the current and past values of itself as well as on current and past values of other relevant quantities. Since there were three temperatures to predict, three distinct models were trained and, so as to avoid augmenting the Gaussian process with unnecessary features, we made use of domain knowledge. Rooms that are not neighbors do not directly influence each other's temperatures; similarly, changing the valve position of AHU~1 has no effect on any temperature besides $T_1$. Initially, all exogenous signals $T_\text{sup}$, $T_\text{out}$ and $R_\text{sol}$ had been included into all models to boost their prediction capabilities. Nevertheless, we later realized that $R_\text{sol}$ was a significant covariate only for $T_3$ as discussed in Section~\ref{sec.ModelTrainingAndTesting}. Field tests unveiled a high correlation between the MPC computation time over the day and the solar radiation curve. Since having predictable rather than fluctuating solve times was a project requirement, we decided not to employ $R_\text{sol}$ as a feature in any GP. The definitive set of employed features is reported in Table~\ref{tab.features}, where the delay parameter $l$ indicates the number of current plus past values used from that particular physical signal. By using the mean functions \eqref{eq.GPmean} to evolve the temperature dynamics, we arrive at the final models
\begin{subequations}
\begin{align}
    T_{1,t+1} &= \mu_{1}(T_{1,t}, T_{1,t-1}, T_{2,t}, \theta_{1,t},T_{\text{sup},t},T_{\text{out},t}) \\
    T_{2,t+1} &= \mu_{2}(T_{1,t}, T_{2,t}, T_{2,t-1}, T_{3,t}, \theta_{2,t},T_{\text{sup},t},T_{\text{out},t}) \\
    T_{3,t+1} &= \mu_{3}(T_{2,t}, T_{3,t}, T_{3,t-1}, \theta_{3,t},T_{\text{sup},t},T_{\text{out},t}) 
\end{align}
\label{eq.GPmodels}
\end{subequations}
Concerning the variances \eqref{eq.GPvar}, we opted for not propagating them forward in time since no closed-form expression exists to accomplish this. Instead, the expression \eqref{eq.GPvar} was evaluated in a point-wise fashion to measure uncertainty. The interested reader is referred to \cite{girard2004approximate,mchutchon2011gaussian} for insightful discussions on the matter.

\subsection{Model training and testing}
\label{sec.ModelTrainingAndTesting}

Data collection was carried out from August to November 2021. The final batch consisted of 22,455 points sampled at $T_\text{samp} = 2\,$ mins and comprised closed-loop operation with PI and rule-based controllers (RBCs), as well as a variety of open-loop excitation signals such as ramps and uniformly random inputs. After examining the obtained curves, we concluded that a control period of 10 mins would be a good compromise between operating the HVAC system effectively and not oversampling the temperatures -- given that our model complexity grows with the size of the dataset, the latter aspect was rather important. The data batch was then downsampled by a factor of 5 times, resulting in 4491 points (748 hours). After that, a meticulous post-processing step was necessary to ensure that unreliable periods were discarded, outliers were detected and filtered, and imputation was performed to fill in certain missing entries. The feature vectors and labels were then created for each of the GPs described in Table~\ref{tab.features}, hence defining their training sets. In our particular case, all variables had similar ranges, thus normalization was not necessary. A critical step was to drop feature vectors that were too close to each other (in a Euclidean norm sense), which not only removed redundant information from the batch, but also improved the numerical stability associated with the kernel matrix \cite{williams2006gaussian}. Finally, the GPs modeling rooms 1, 2 and 3 had respectively 231, 308 and 281 points, which correspond to approximately 38, 51 and 47 hours worth of data. Since these sets did not come from a single experiment, but are an informative subset of the 748 hours initially available to us, they provided enough prediction capabilities to our non-parametric models.

\begin{figure*}[!t]
    \centering
    \includegraphics[scale=0.29]{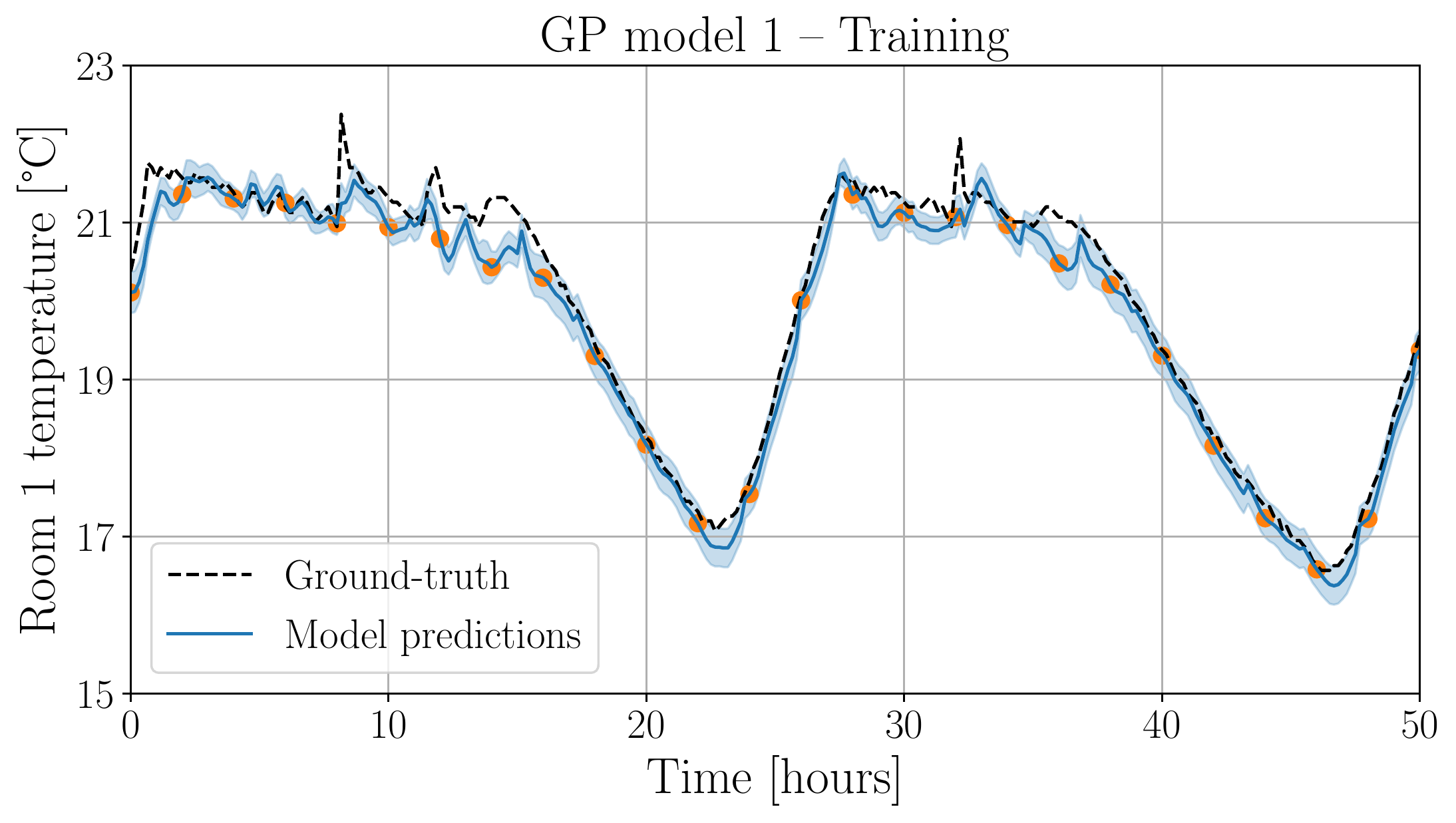}
    \includegraphics[scale=0.29]{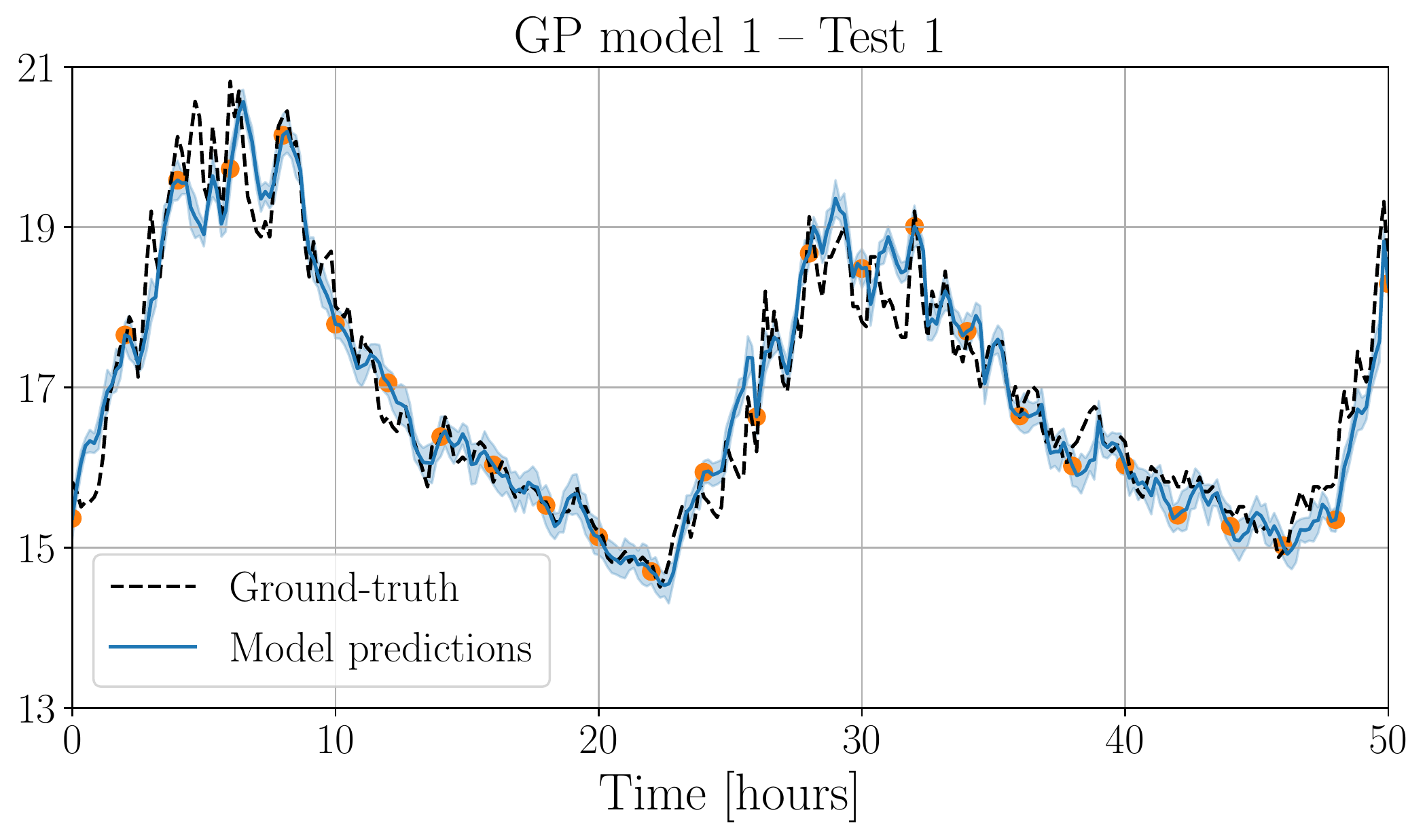}
    \includegraphics[scale=0.29]{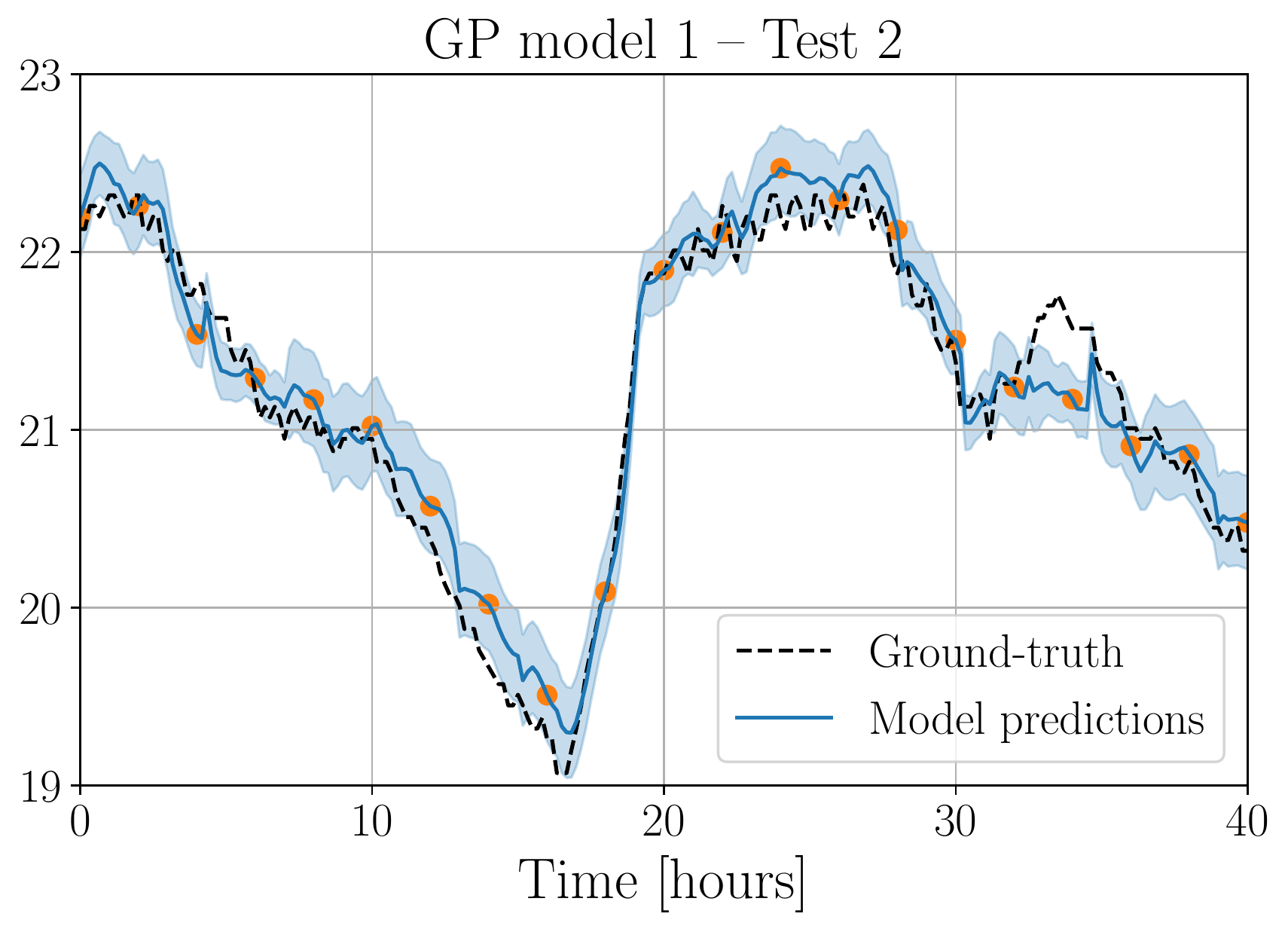} \\[4pt]
    \includegraphics[scale=0.29]{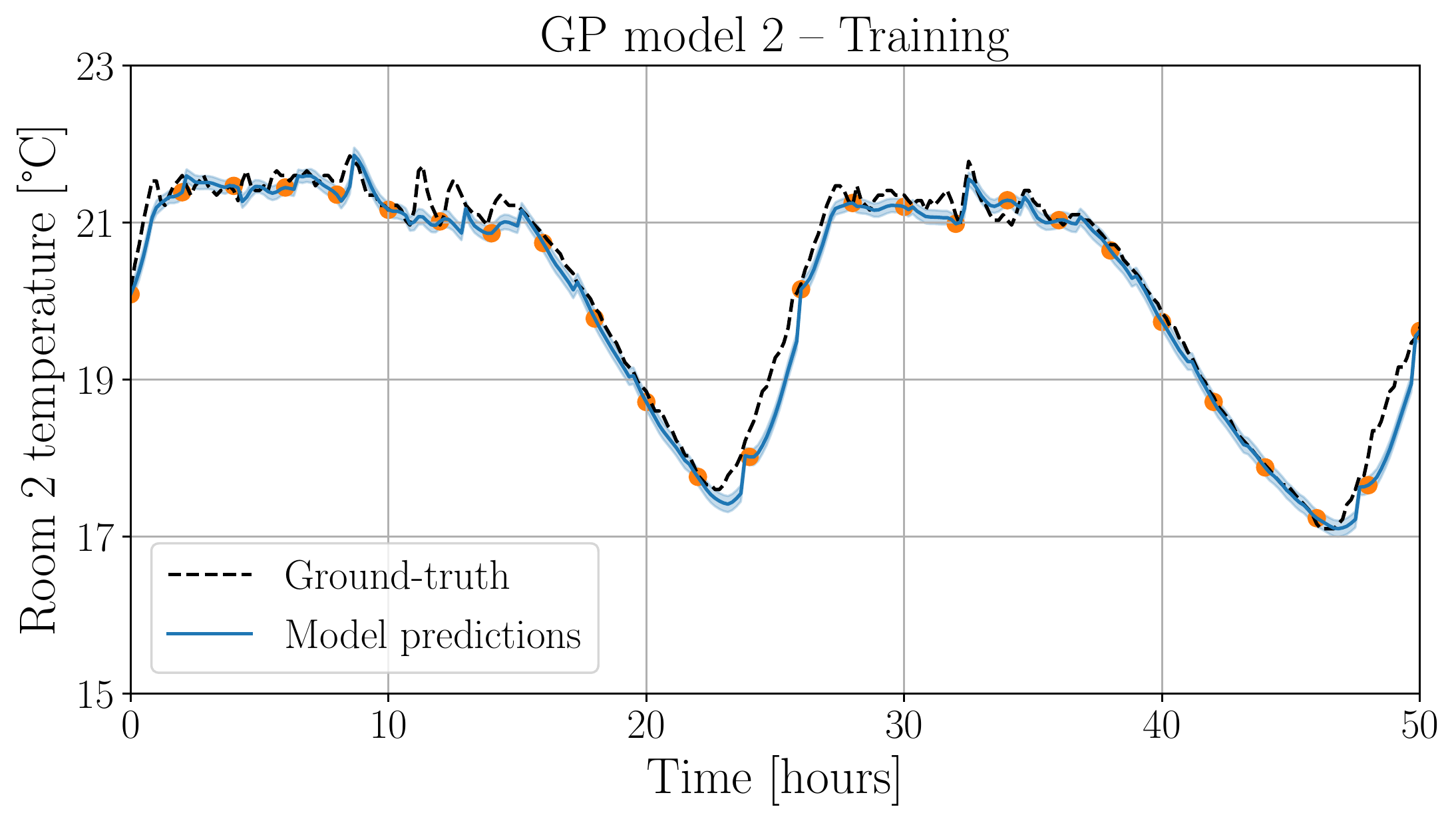}
    \includegraphics[scale=0.29]{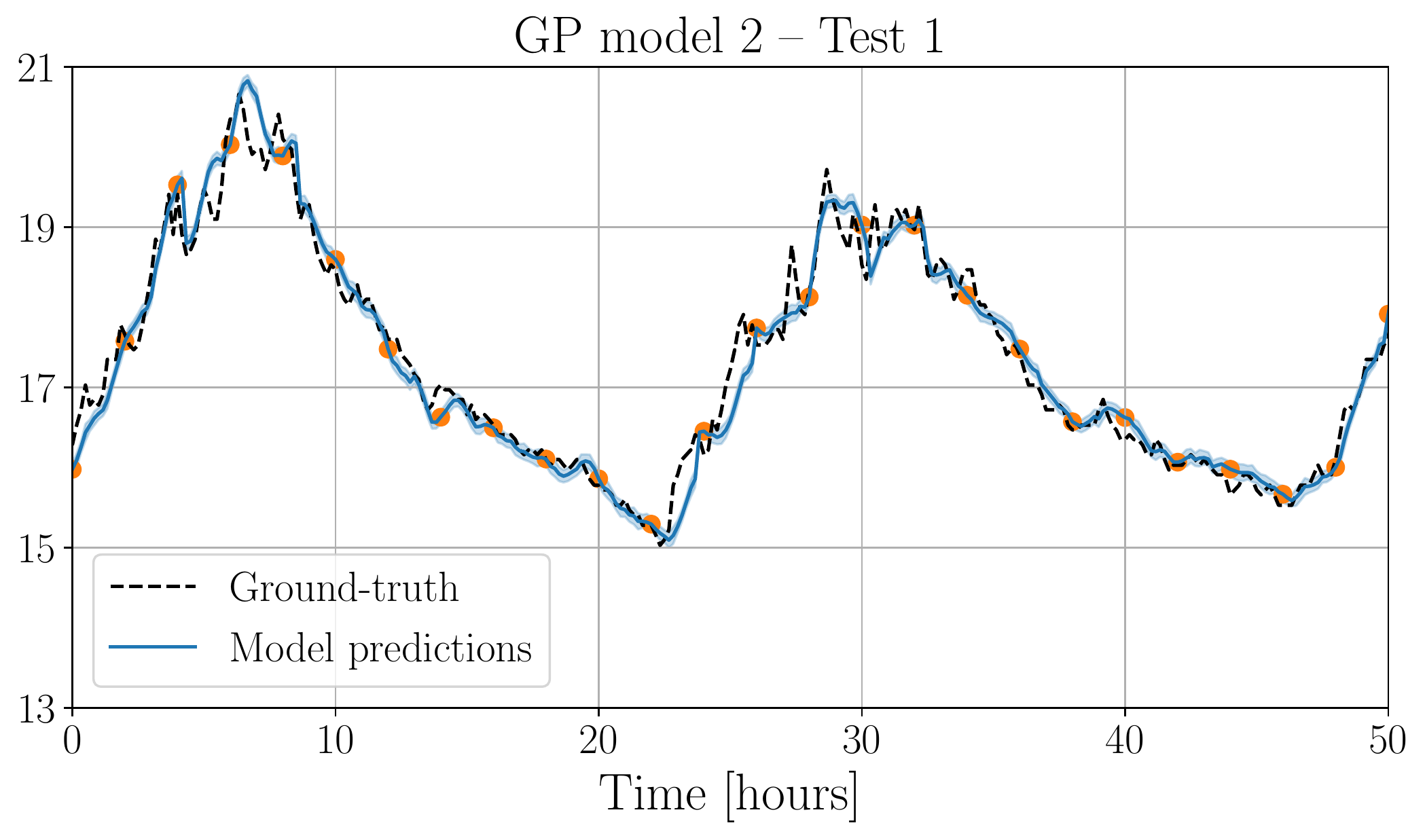}
    \includegraphics[scale=0.29]{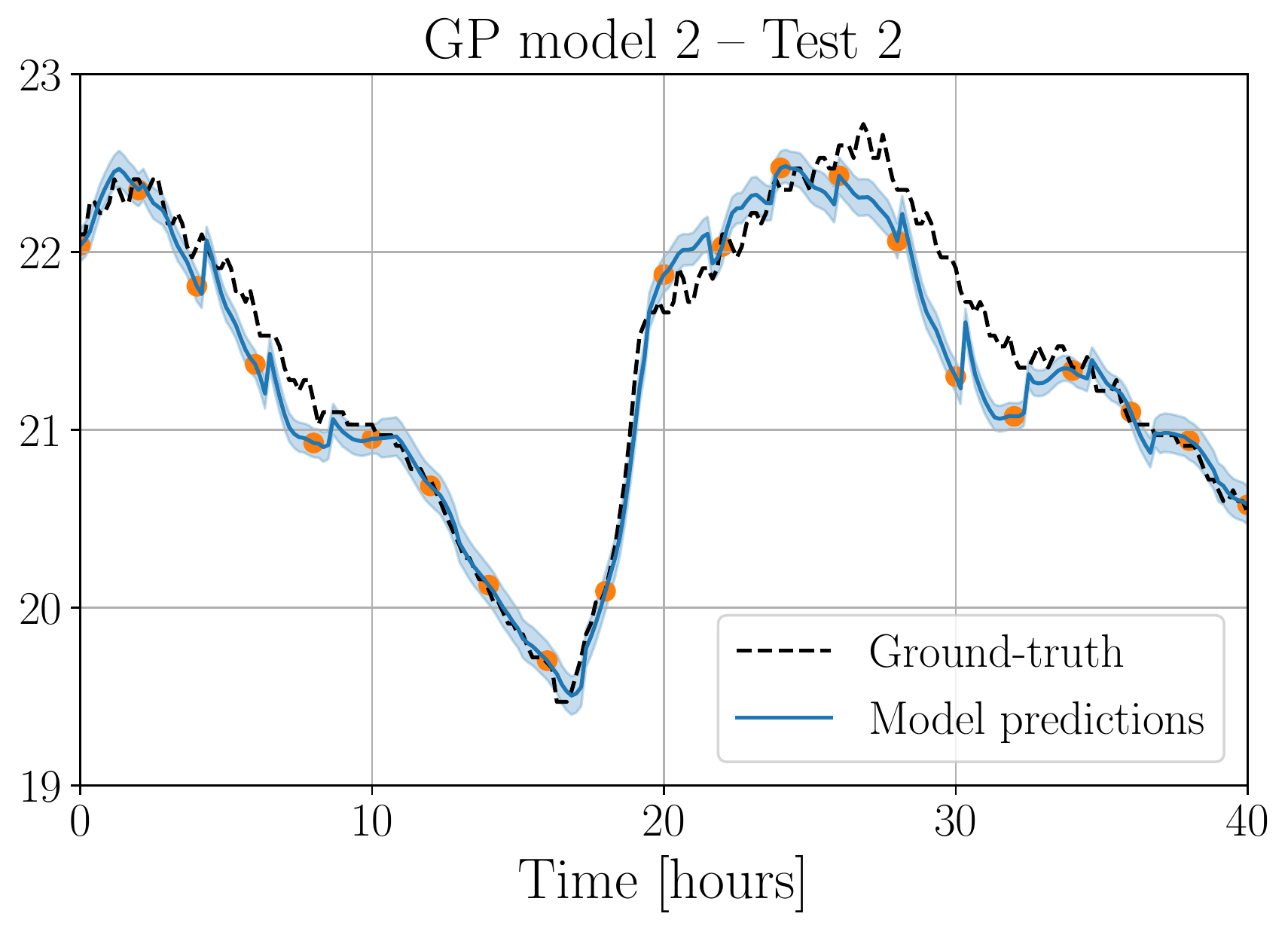} \\[4pt]
    \includegraphics[scale=0.29]{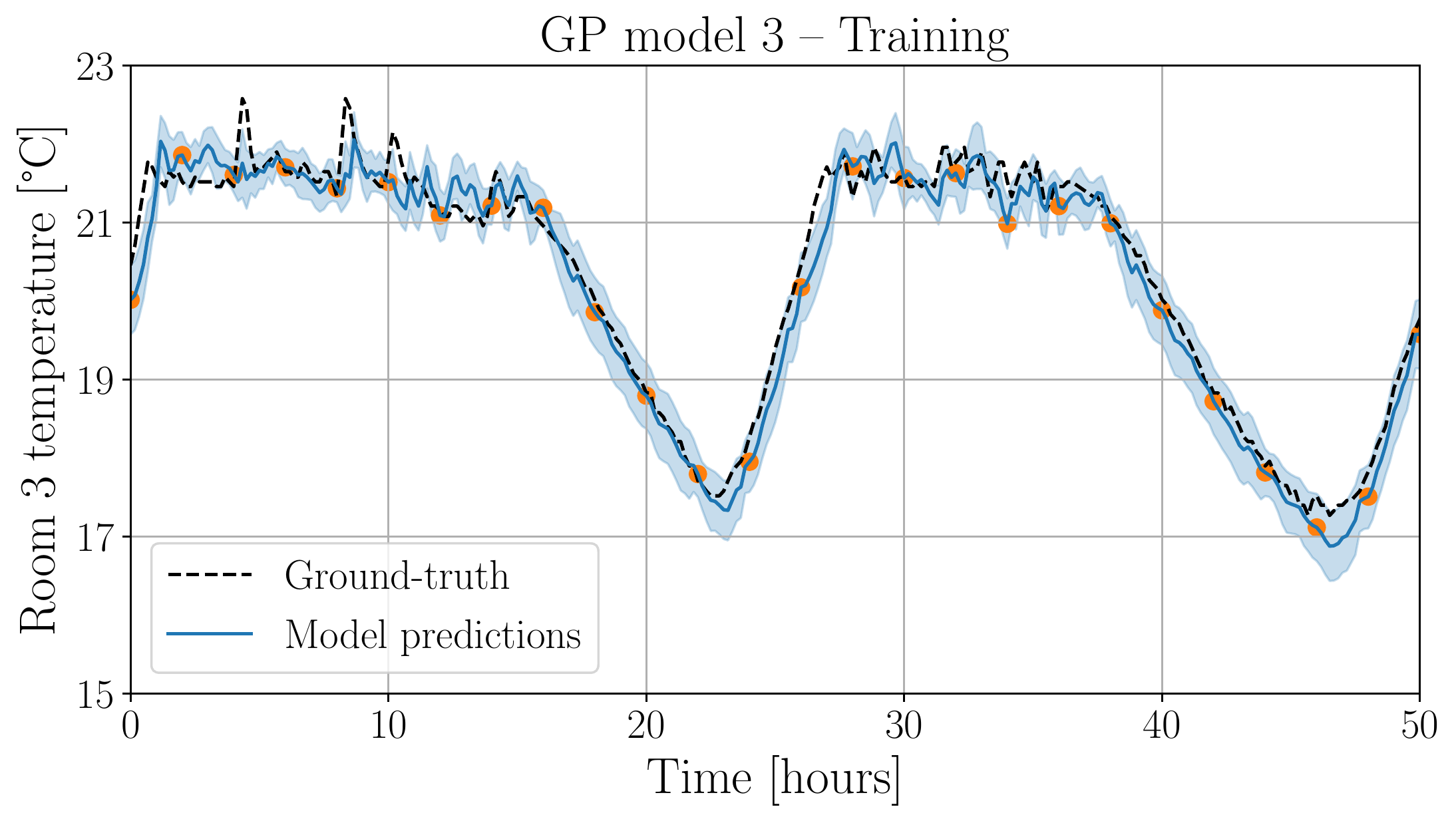}
    \includegraphics[scale=0.29]{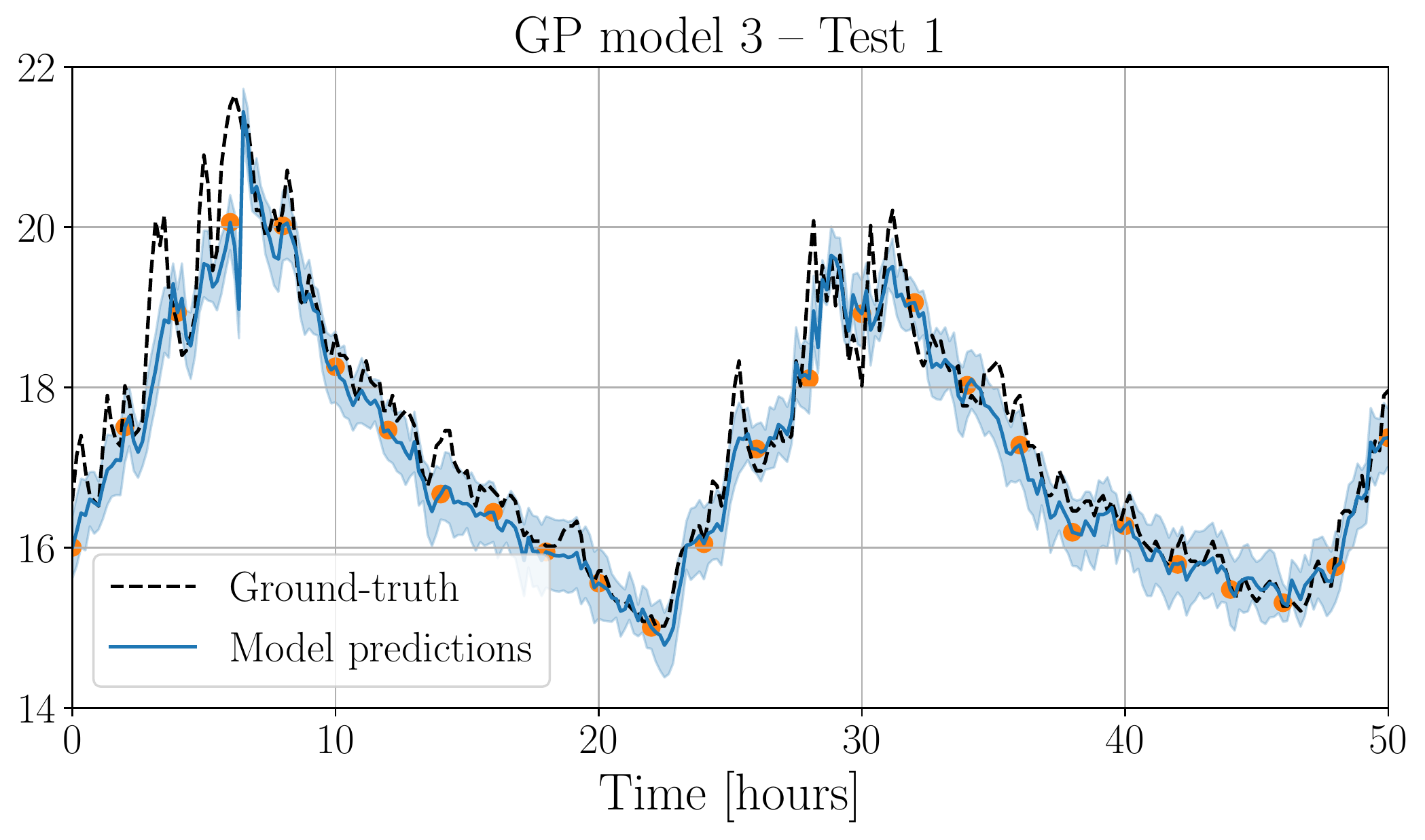}
    \includegraphics[scale=0.29]{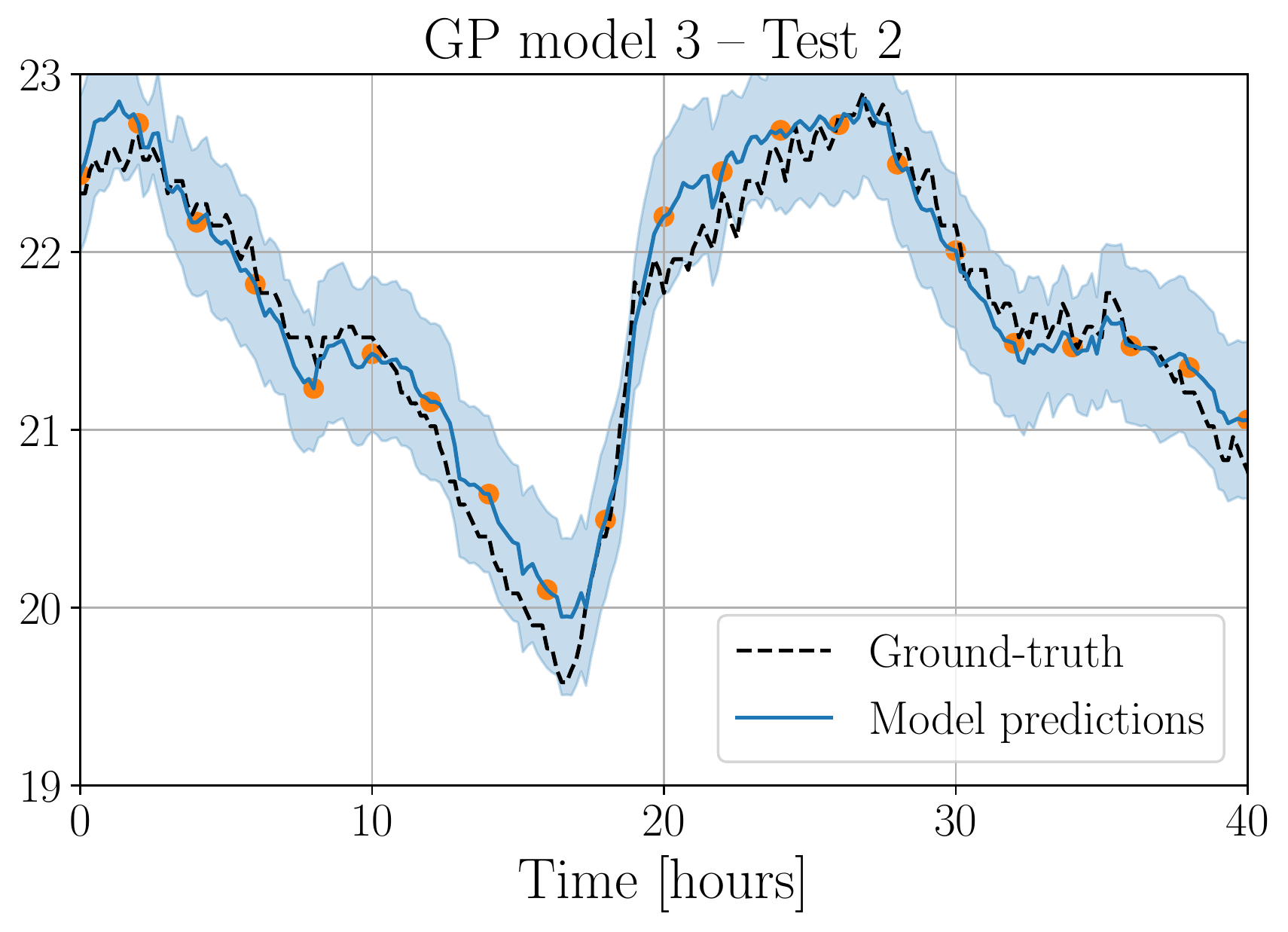}
    \caption{Training (left) and test results (center, right) of the Gaussian process models based on real experimental data. The mean predictions are shown in solid blue, whereas the uncertainty envelope of two standard deviations, in light blue. The sampling period is 10 minutes. The mean predictions were obtained by simulating the GP auto-regressive models forward in time for 2 hours and only feeding back new temperature information at the orange dots. The training plots only show a portion of the training set.}
    \label{fig.trainTestRes}
\end{figure*}

All GPs were defined and trained with the aid of the \texttt{GPflow2} \cite{gpflow} and \texttt{SciPy} \cite{SciPy} packages. No priors were placed on the hyperparameters and we employed the marginal likelihood criterion along with the limited-memory BFGS optimization algorithm to tune them. Training the models with the aforementioned number of points took consistently less than 5 seconds each on a 2.4 GHz, core i9-9980HK machine. The obtained training and test results can be seen in Figure~\ref{fig.trainTestRes}. We highlight that the plots show multi-step ahead predictions over a horizon of 2 hours, that is, 12 time steps, correcting for the temperature mismatch only at the orange points. Assessing the prediction quality of the models in this way was necessary as they were to be used within an MPC formulation. It is worth noting that, even though the left plots are labeled as ``training results'', the models incorporated only a small fraction of those features due to our dropping of nearby data-points. The central and right-side plots show the predictions over a period of 50 and 40 hours, but in completely new scenarios, never presented to the model during training. 

Inspecting Figure~\ref{fig.trainTestRes}, we see that the mean predictions mostly follow the underlying ground-truth signal during training. The disparity among the rooms is in their uncertainty bands: whereas model 1 and 2 presented moderate levels of spread, model 3 showed a fairly large one. We believe this uncertainty to stem from room~3 being the most exposed one in terms of direct solar radiation, and from $R_\text{sol}$ not being a feature of its GP. We remind the reader that $R_\text{sol}$ was disregarded to accelerate the real-time computations and ensure that the optimization problem was solved within the time allocated to it. By taking this larger uncertainty into account, we were able to avoid violating constraints when closing the loop with the MPC controller. The outcome of the test phase was qualitatively similar to the training results, aside from some additional performance degradation close to the high temperature peaks. Overall, we deemed the results reasonable given the challenging two-hour horizon of the prediction task.

\subsection{Learning the chiller energy consumption}
\label{sec:learningChiller}

\begin{figure*}[!t]
    \centering
    \includegraphics[width=0.35\linewidth]{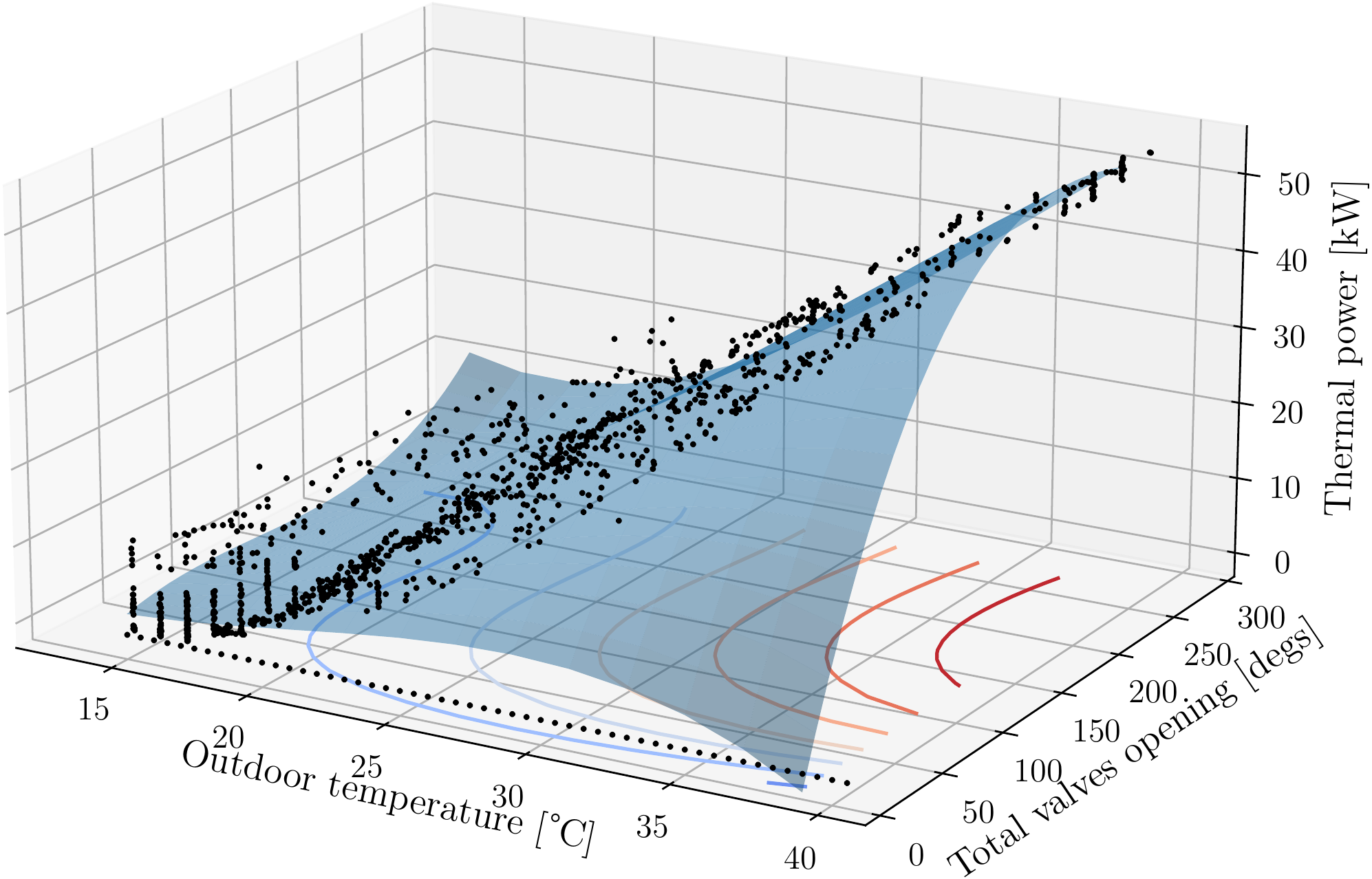} 
    \hspace{10pt}\includegraphics[width=0.26\linewidth]{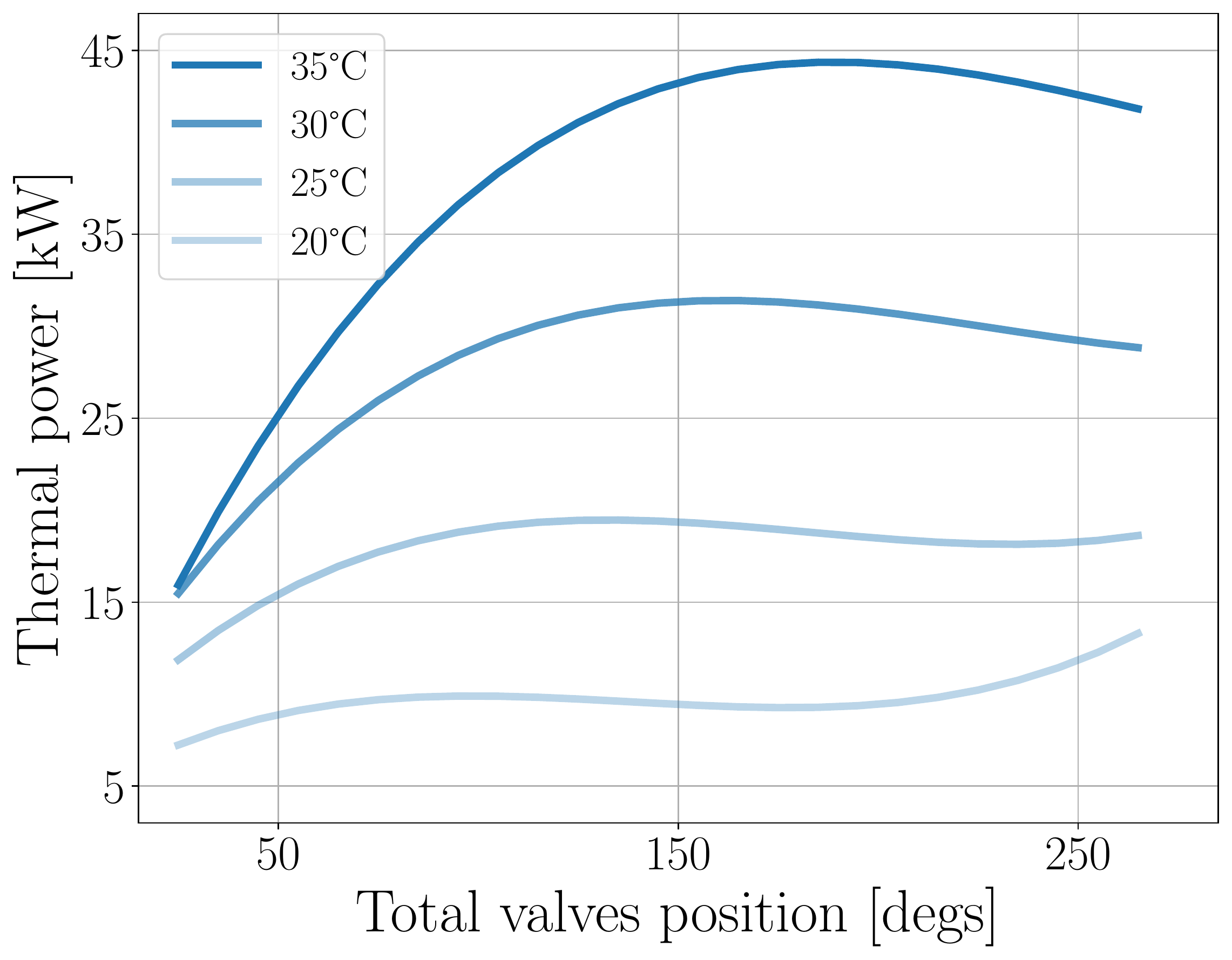}
    \includegraphics[width=0.26\linewidth]{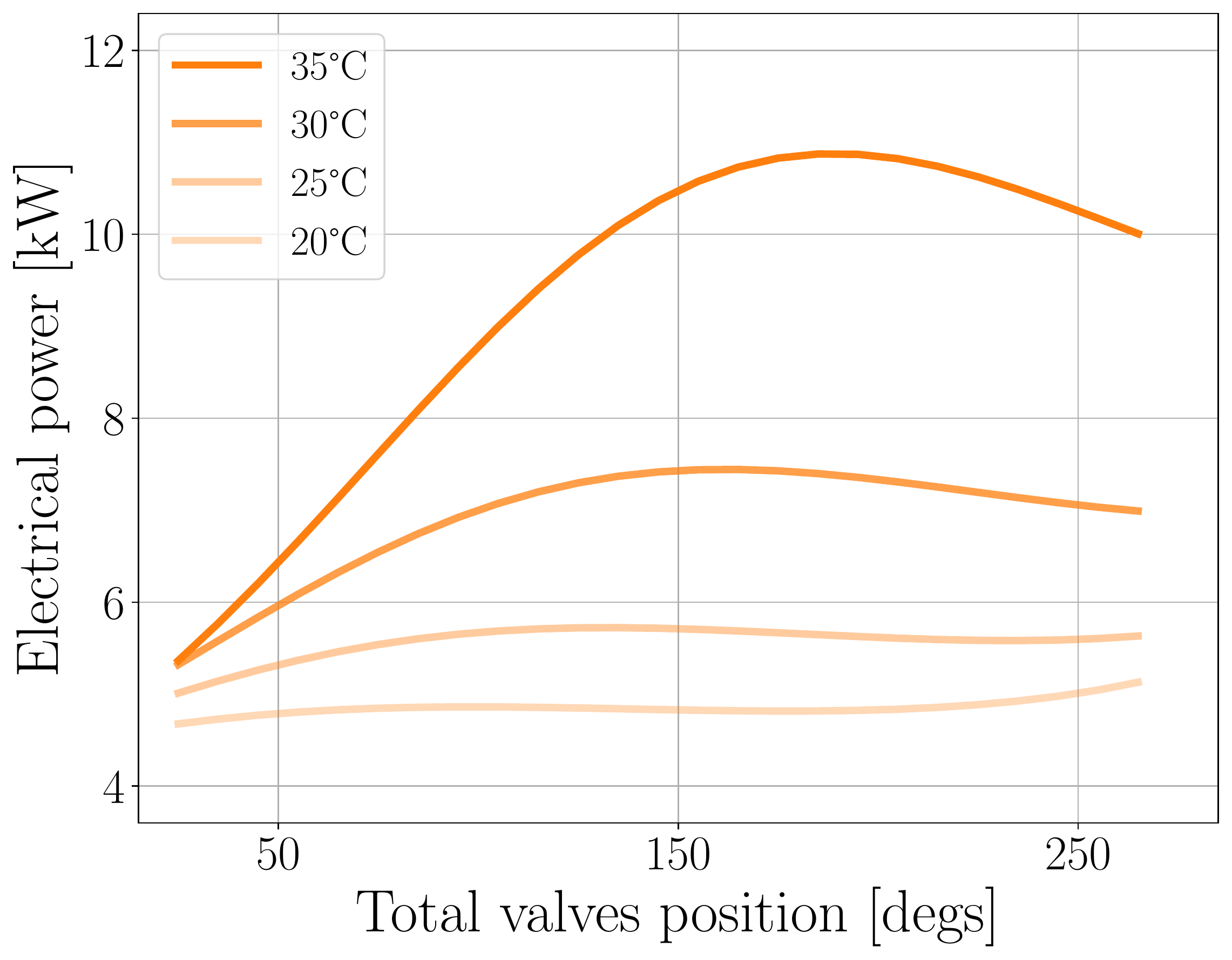}
    \caption{(Left) Reconstruction of the chiller refrigeration surface. Data were collected during a period of 203 hours, including both of open-loop excitation as well as closed-loop operation. (Center) Thermal power $Q(T_\text{out},\Theta)$ and (right) electrical power $E(T_\text{out},\Theta)$ curves of the chiller as a function of the valves openings $\Theta$. The plots consider typical outdoor temperature values $T_{\text{out}} = 20, 25, 30$ and $35^{\circ}$C.}
    \label{fig.thermalAndElectrical}
\end{figure*}

We tackled the problem of operating the system while minimizing its electrical demand with the two-step approach described next. The first goal was to reconstruct the chiller refrigeration curve from historical data, more specifically, from the volumetric air-flow rates along with the outdoor temperature and the supplied air temperatures. Based on these quantities, the thermal power delivered by the chiller was inferred. Next, we used as features the outdoor temperature $T_\text{out}$ and the sum of the valve positions $\Theta = \theta_1 + \theta_2 + \theta_3$, the latter correlating with the water flow through the AHU coils (see Section~\ref{sec.controlProb}). A representative dataset was gathered over a period of 203 hours, which encompassed both random open-loop excitation and closed-loop operation. During such period, the AHU valves ranged from being completely open to being fully closed, and the ambient temperature varied from 15 to 40 degrees Celsius. The data distribution can be seen in Figure~\ref{fig.thermalAndElectrical}. We remark that the portion of the domain where the outdoor temperature is high and the total valve openings are low is not populated with samples due to operational constraints of the system, a common issue in HVAC control \cite{maddalena2020data}. The last step was to augment the batch with a grid of $T_\text{out}$ values paired with $\Theta=0\,$degs and $0\,$kW labels to represent the zero water-flow regime.

Polynomial ridge regression was performed to fit the data described above. After experimenting with different model orders, we found that a cubic model provided a good balance between describing the observed points and not overfitting them. The results are presented in Figure~\ref{fig.thermalAndElectrical}, where the obtained model analytical expression is
\begin{align}
    &\text{Q}(T_\text{out},\Theta) = -3.15 \, T_\text{out} - 3.03 \,\texttt{e}^{-2} \, \Theta + 1.73\,\texttt{e}^{-1} \, T_\text{out}^2 \nonumber \\
    &- 1.56\,\texttt{e}^{-3} T_\text{out} \Theta + 3.09\,\texttt{e}^{-4} \Theta^2 - 2.75\,\texttt{e}^{-3} T_\text{out}^3 \label{eq.thermalEnergy} \\
    &+ 4.90 \,\texttt{e}^{-4} T_\text{out}^2 \Theta - 6.86 \,\texttt{e}^{-5} T_\text{out} \Theta^2 + 2.56 \,\texttt{e}^{-6}\Theta^3 + 20.22 \nonumber
\end{align}
with \texttt{e}$^{n}$ being a shorthand for $\times10^{n}$. As a second step, we fit a concave coefficient of performance (COP) curve, which is typical for variable-speed compressor chillers and dictates how efficient they are in converting electrical to thermal energy. The final utilized COP curve was
\begin{equation}
\begin{split}
    \text{COP}(Q) = 3.30 \, \texttt{e}^{-7} Q^4 - 2.69 \, \texttt{e}^{-5} Q^3 - 2.67 \, \texttt{e}^{-3} Q^2 \\ + 2.34 \, \texttt{e}^{-1} Q^1 - 4.45 \, \texttt{e}^{-4}
    \label{eq.COP}
\end{split}
\end{equation}
which, given the thermal range displayed in Figure~\ref{fig.thermalAndElectrical}, implies in a performance coefficient varying approximately between 1.5 and 4.5.

With the thermal model \eqref{eq.thermalEnergy} and the COP curve \eqref{eq.COP} at hand, the electrical power could be calculated according to $E = Q(T_\text{out},\Theta)/\text{COP}(Q(T_\text{out},\Theta))$, measured in kW. Several slices of the thermal power surface, and their associated electrical power counterparts are presented in Figure~\ref{fig.thermalAndElectrical}. The plots illustrate how the curves change depending on the outdoor temperature, and how strongly the electrical power profile is affected by this external factor. In particular, one notices that when the outside temperature is high, it is more economic to open the valves and increase the chilled water flow rather than keeping them partially closed. Clearly though, the real-time optimal position for them will depend on the system dynamics, the desired temperature envelope and the external disturbances.

\subsection{MPC formulation and numerical computations}

Given the learned GP models $\mu_i$, $i=1,2,3$ described in \eqref{eq.GPmodels}, a given maximum temperature $T_{\max}$, and our reconstructed electrical power surface, we formulate the following optimization problem to control the valves $\theta_i$ while reducing the chiller energy consumption $E_t$
\begin{subequations}
\begin{align}
    \min & \quad \sum_{t=0}^{N-1}  \left(E_t + \rho \Delta_t \right) + \rho_N \Delta_N\\
    \text{s.t.}
    & \quad T_{t+1} = \mu(T_t,\theta_t,T_{\text{sup}},T_{\text{out}}) \label{eq.constr1} \\
    & \quad T_t + \beta \, \text{var}^{1/2} (T_t,\theta_t,T_{\text{sup}},T_{\text{out}}) \leq T_\text{max} + \delta_t \label{eq.constr2} \\
    & \quad E_t = Q(T_\text{out},\Theta_t) / \, \text{COP}(Q(T_\text{out},\Theta_t)) \label{eq.constr3} \\
    & \quad \theta_{\text{min}} \leq \theta_t \leq \theta_{\text{max}} \label{eq.constr4} \\
    & \quad \delta_t \geq 0 \label{eq.constr5}
\end{align}
\label{eq.optControlProb}%
\end{subequations}
where $\Theta_t = \sum_{i=1}^{3} \theta_{i,t}$ is the sum of all valve positions. The variables $\delta_t$ in \eqref{eq.constr2} are positive slacks introduced to avoid infeasibility. If needed, these can relax the temperature constraint so that the solver can return a viable control plan. Of course, their use is heavily penalized in the objective, where $\Delta_t = \sum_{i=1}^{3} \delta_{i,t}^2$ and $\rho$, $\rho_N$ are large constants, which in our case were respectively set to $100$ and $200$. The temperature constraint \eqref{eq.constr2} also accounts for prediction uncertainty as it includes the standard deviation $\text{var}^{1/2}$. Its use confers on the formulation a risk-aware quality and robustifies the closed-loop operation. The degree of conservativeness is controlled by the constant $\beta$, chosen to be $2$ as in Figure~\ref{fig.trainTestRes}. The prediction horizon was set to $N=12$ steps, which translates to 2 hours. As suggested by our notation, $T_\text{out}$ and $T_\text{sup}$ were kept constant throughout all prediction steps--but updated from one sampling period to the next. Finally, our maximum temperature value was $T_{\max} = 21$ degrees Celsius.

The optimization problem \eqref{eq.optControlProb} was written in \texttt{Python} with the aid of CasADi \cite{casadi}, an automatic-differentiation package that provides gradient information for numerical solvers--in our case, the interior-point method IPOPT. As is customary in predictive control, \eqref{eq.optControlProb} was recursively solved on-line with the most recently available system information, with only the first optimal control action being transmitted to the valves. We underline that the main source of complexity in \eqref{eq.optControlProb} is the presence of the constraints \eqref{eq.constr1} and \eqref{eq.constr2}, which are highly non-linear due the GP mean and variance. Since convexity is absent, multiple local optima might exist, a fact that was indeed verified in practice. By intelligently providing solvers with high-quality initial guesses, this problem can be mostly overcome. Our particular case study relied on initializing the numerical solver with control, temperature, slack and energy trajectories obtained with a virtual PI controller. The intuition was to allow the MPC loop to build on such an initial guess and further optimize operation. For a detailed study on solve times and how the number of GP data-points impacted them, see Appendix~A.

\section{Simulation and Experimental Results}
\label{sec.SimExpRes}

\subsection{Field tests and results}
\label{sec.expRes}

\begin{figure*}[!t]
    \centering
    \includegraphics[width=0.85\linewidth]{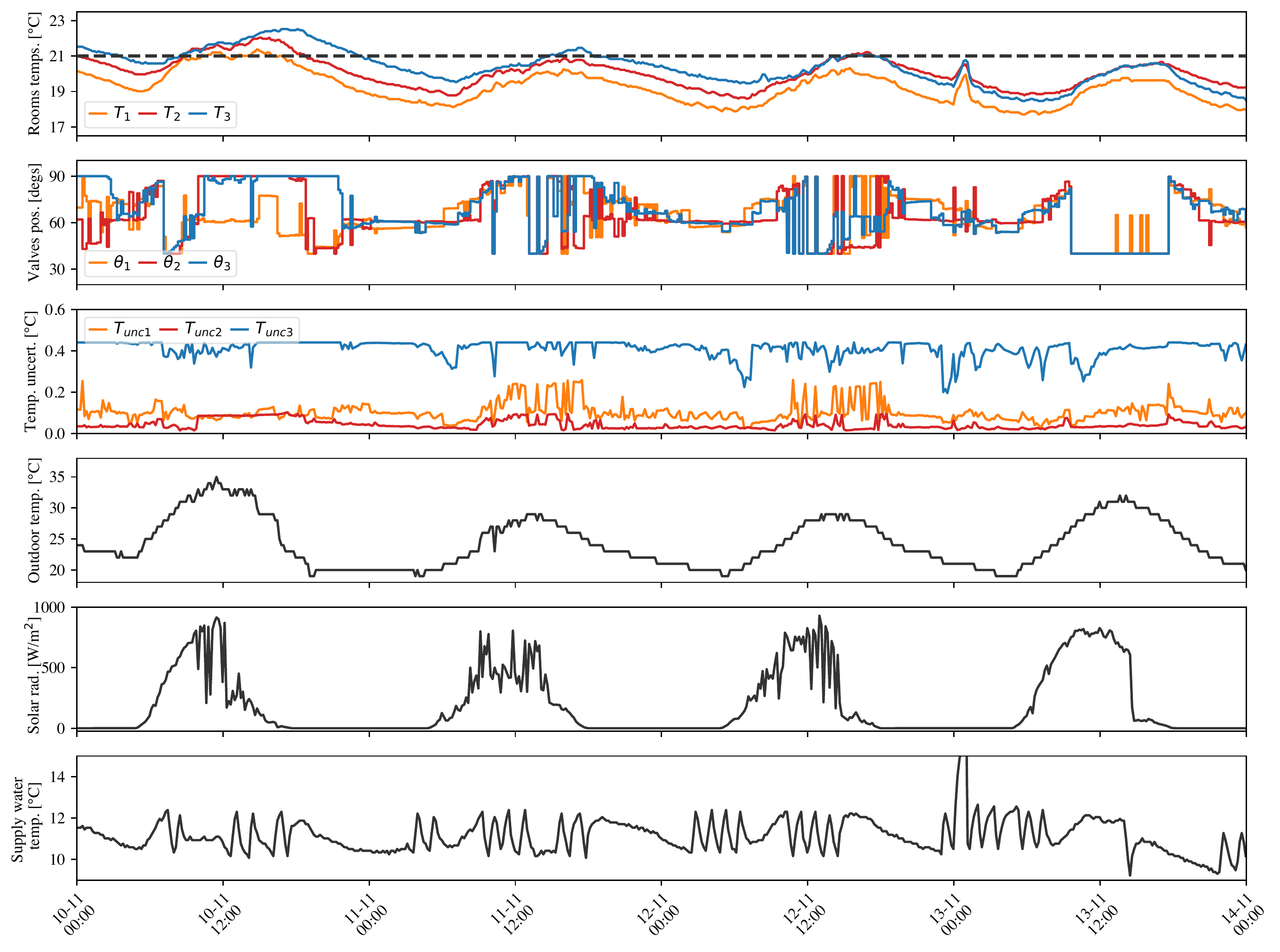} 
    \caption{MPC experimental results over four days: indoor temperature, valve position and uncertainty estimate associated with each room (top three plots); outdoor temperature, solar radiation and AHUs supply water temperature (bottom three plots). The system was sampled and controlled with a periodicity of 10 minutes.}
    \label{fig.prettyCool}
\end{figure*}

The previously described Gaussian process-based MPC formulation was deployed on the local computer and used to operate the HVAC system during multiple days in the months of October and November 2021. We report in Figure~\ref{fig.prettyCool} a four-day uninterrupted experiment carried out from November 10 to November 13 that is rather representative of the local internal and external conditions. The plots show the room temperatures and the ``immediate'' uncertainty associated with the GP predictions: $T_\text{unc} = \beta \text{var}^{1/2}$ as employed in the formulation \eqref{eq.constr2}, and evaluated for the next time-step. Both outdoor signals, the temperature and the solar radiation, are also given. The reader is reminded that, although the latter contributes with additional heat gains, it is completely unknown to the controller as explained in Section~\ref{sec.ModelTrainingAndTesting}. We highlight that the curves displayed in the figure were not filtered in any way; the sole manipulation performed with the data was the imputation of the missing temperature entries using linear interpolation. These points, however, accounted for only 43 out of the 1728 indoor temperature values gathered during the four-day experiment.

Consider first the day November 10 and note the relatively high internal room temperatures when the experiment started, which were the consequence of a harsh previous day. The MPC controller used some control authority to bring the temperatures below the 21-degree line and then partially closed the valves. After the morning shift started (7 am), even though $\theta_2$ and $\theta_3$ were fully open, $T_2$ and $T_3$ violated the constraints and were only brought below 21 degrees late that evening. High initial conditions along with a peak outdoor temperature of 35 degrees overloaded the cooling system, causing violations of the indoor temperature constraint in two rooms. 

The two days that followed (November 12 and 13) were less warm and, as a result, the MPC controller successfully modulated the valves so as to guarantee constraint satisfaction. It is evident how $\theta_1$, $\theta_2$ and $\theta_3$ assume lower values when $T_\text{out}$ is low, and tend to saturate at their maximum during working hours, which matches our intuition. 

Lastly, we focus on the data from November 13, where one can readily see a sudden peak in the indoor temperatures, being also present in $T_{\text{sup}}$. This was caused by a momentary halt in the water pumps responsible for the chilled water circuit--an event that could be regarded as a fault from a control system perspective. During this period, as there was no water circulation through the AHU cooling coils, there was also no refrigeration and the indoor spaces received warm air since the fans were kept on. As soon as the pumps were again activated, the chiller immediately decreased the supply water temperature and the operation was normalized. During daytime, the indoor climate was kept within the desired limits despite the valves staying saturated at their low values, even at noon. The fact that almost no additional actuation was needed is due to that day being a Saturday, when no operations are scheduled and the three doors present in the environment are minimally opened and closed. This demonstrates how strong the internal heat gains and unmeasured disturbances normally are.

\subsection{A comparison among different algorithms}

\begin{figure*}[!t]
    \hspace{0pt}
    \includegraphics[width=0.95\linewidth]{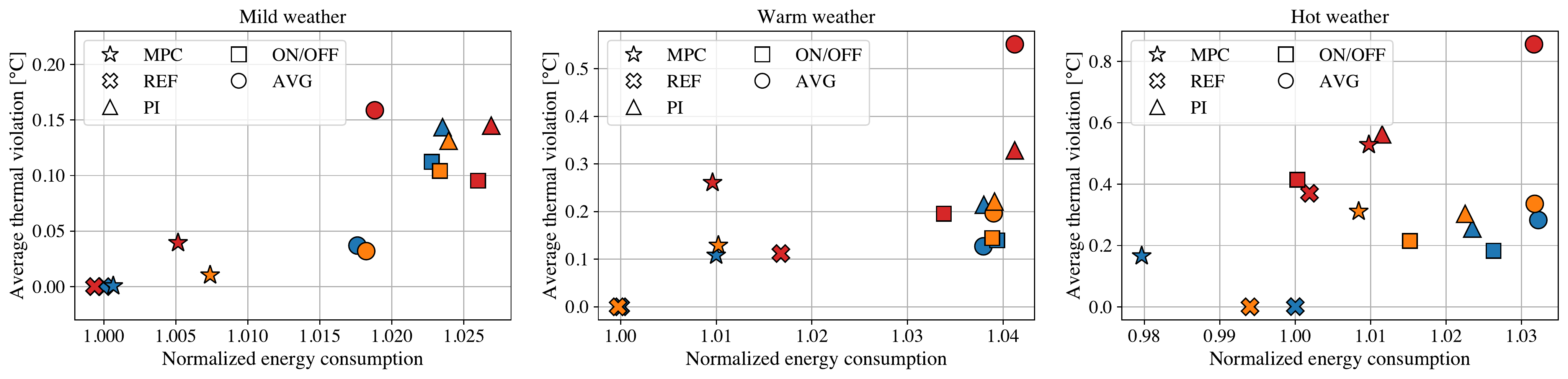} 
    \caption{Simulation results of the normalized energy consumption and thermal performance (average temperature bound violation) of different control strategies. Three weather profiles were considered: mild, warm and hot. The indoor temperatures were initialized at different values according to the color scheme: {\Large \textcolor{color_20}{\textbf{-}}} 17 °C, {\Large \textcolor{color_21}{\textbf{-}}} 19 °C, {\Large \textcolor{color_22}{\textbf{-}}} 21 °C.}
    \label{fig.powerRes}
\end{figure*}

To assess the efficiency gains as well as the thermal performance of the deployed strategy, MPC was compared to alternative algorithms, all subject to exactly the same environmental conditions by means of simulations. We underline that this simulation model was calibrated on data that was \textit{not} included in the GPs training set, thus putting to test the MPC prediction capabilities. The disturbance signals $T_\text{out}$, $T_\text{sup}$ and $R_\text{sol}$ from November 10 were employed, and the indoor temperatures of the three rooms were uniformly initialized at values ranging from $17$ to $21$ degrees Celsius. The outdoor temperature profile was processed to yield three different weather scenarios: hot weather, which was exactly the same $T_\text{out}$ curve seen in Figure~\ref{fig.prettyCool}; warm weather, a $-2$°C shifted version of it, peaking at 33°C around noon; and mild weather, a $-5$°C shifted version of it, peaking at peaking at 30°C. Besides the MPC algorithm \eqref{eq.optControlProb}, the following were also tested:
\begin{itemize}
    \item An MPC controller (herein referred to as REF) with perfect prediction capabilities, perfect disturbance information ($T_\text{out}$ and $T_\text{sup}$) and a long prediction horizon of five hours.
    \item PI controllers featuring anti-windup schemes and feedforward components to enhance their performance.
    \item Rule-based ON/OFF controllers that set the valves respectively to $\theta_\text{min}$ and $\theta_\text{max}$ if the indoor temperatures were below or above the set-point. 
    \item An average $(\theta_\text{max}-\theta_\text{min})/2$ controller (AVG) whose instantaneous values are selected by sampling the interval $\theta_\text{min}$ to $\theta_\text{max}$ using a uniform distribution.
\end{itemize}

In order to gauge the energy saving potential of the HVAC plant, we executed the REF strategy described above. This algorithm can reach significant efficiency gains while guaranteeing thermal comfort since it exploits a perfect internal model as well as perfect forecasts of the outdoor and supply water temperatures.

The obtained normalized energy consumption results and average room thermal comfort violations are shown in Figure~\ref{fig.powerRes}. As the chiller curves described in Section~\ref{sec:learningChiller} were employed and non-linear controllers were tested, the reader is reminded that there is a non-trivial relationship between the weather conditions and indoor temperatures, and the resulting energy consumption. Lastly, in every individual plot, the corresponding REF $17\,$°C energy consumption was used as a normalization factor.

Glancing at the mild and warm weather plots, one notices how the REF and MPC data tended to be close together, and relatively far from the PI, ON/OFF and AVG clusters. Moreover, the REF and MPC points were also mostly to the left side and vertically below the other data given the same indoor temperature conditions--thus confirming their superior performance in terms of energy efficiency and indoor climate regulation. The ON/OFF and PI controllers yielded overall similar numerical results and, surprisingly, were outperformed by the AVG scheme under mild weather and starting indoor temperatures of $17\,$°C and $19\,$°C. AVG nevertheless performed poorly under warm weather and $21\,$°C, and hot weather in general.

By analyzing the horizontal scales and contrasting REF to PI, ON/OFF and AVG, one concludes that this particular HVAC plant could have its efficiency boosted by approximately 2.5\%, 4\% and 5\% respectively in the mild, warm and hot weather scenarios. Notice how, as opposed to studies that minimize the cumulative thermal energy such as \cite{bunning2020experimental}, our numbers refer to the electrical energy consumed by a chiller and expressed through a highly non-convex function. Moreover, as the hospital was not subject to time-varying electricity prices, the contrast among control strategies was not as broad as for instance the one reported in \cite{joe2022investigation}. The proposed MPC strategy \eqref{eq.optControlProb} attained results close to the aforementioned maximum percentages. Quantitatively, MPC lead to a maximum energy efficiency improvement of 2.29\%, 3.13\% and 4.76\% respectively in mild, warm and hot weather, when compared to the PI and ON/OFF counterparts.


\section{Final Remarks and Conclusions}

Our study has demonstrated the suitability of Gaussian process to learn buildings dynamics featuring forced-air HVAC systems. Aside from feature selection, which exploited information that is easy to infer from the building geometry, training the models was straightforward and did not require informed initial guesses for the hyperparameters. The most time-consuming step in the pipeline was certainly data acquisition and cleaning, during which spurious periods were identified and dropped from the dataset. Nevertheless, the same procedure would certainly also have to be carried out if one were to craft linear models for instance. The inherent robustness to outliers of one or another model class and associated training objectives could be the matter of future investigations. Overall, we regard GPs as being an appealing modelling framework for the building community that has potential to deliver good predictive performance as showcased in Section~\ref{sec.ModelTrainingAndTesting}.

Combining GPs and MPC results in non-convex optimization problems and relatively high solve times. The analysis carried out in Appendix~A along with the experimental results reported in Section~\ref{sec.expRes} show how this combination is still amenable to real-time implementation in the domain of building temperature control. Our methodology was capable of regulating the indoor temperature whenever enough actuation power was available despite the strong internal heat gains. Moreover, its performance closely approached the HVAC plant limits as estimated by us and lead to electrical energy savings of up to 4.76\% when compared to widespread control techniques.

We envision future investigations exploring two distinct paths. Firstly, speeding up computations to allow for longer prediction horizons and, thus, better economical performance. Seeing that all computations were carried out in a centralized fashion in this study, reducing the solve times would be possible in principle by exploiting algorithm parallelization. Secondly, moving toward higher-level problems such as coordinating multiple buildings that operate under time-varying electricity prices or participate in demand response programs.


\section*{Acknowledgments}

E. T. M. and C. N. J. received financial support from the Swiss National Science Foundation under the RISK project (Risk Aware Data-Driven Demand Response, grant number 200021 175627). The authors express their gratitude to the São Julião hospital administrators and to the local maintenance team for their support.

\appendix
\section{MPC Computation Times}
\label{app.compTimes}

In order to shed light on how the number of training points affects the solve times of the non-convex optimization problem \eqref{eq.optControlProb}, the following study was conducted. Five sets of GP models were trained on distinct datasets with cardinalities $N = 352$, $794$ (precisely the one used in the experiments), $1280$, $1910$ and $2643$. In all scenarios, the total number of points present in each of the GPs was approximately a third of the total number $N$, so that the models were balanced. We then generated random initial conditions, uniformly sampled from sensible intervals: $16 \leq T_{1,2,3} \leq 23$, $9 \leq T_\text{sup} \leq 13$, and $15 \leq T_\text{out} \leq 35$. Finally, we solved the warm-started non-convex MPC \eqref{eq.optControlProb} on a 2.4 GHz, i9 machine $50$ times per scenario and recorded their run times. 

\begin{figure}[!t]
    \hspace{6pt}
    \includegraphics[scale=0.58]{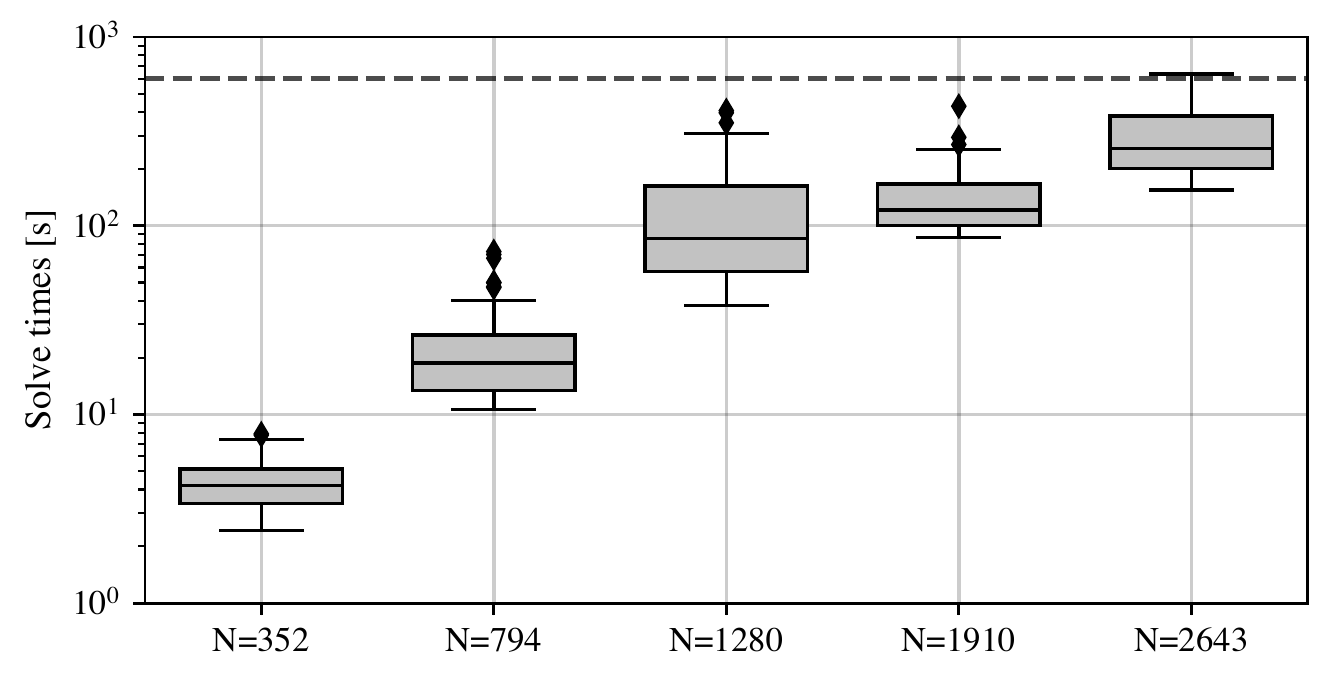}
    \caption{Box and whisker plots of the MPC solve times considering different dataset sizes. The dashed gray line marks our sampling period of $10\,$mins. Each boxplot is based on 50 time samples, obtained using randomized initial conditions.}
    \label{fig.runTimes}
\end{figure}

The results are presented in Figure~\ref{fig.runTimes}, where the vertical scale is logarithmic. The median values of the box plots rose from $4.19$ to $18.69$, $85.42$, $121.12$ and $255.68$ seconds respectively from the smallest to the largest dataset. Although using $N=1910$ points does not seem unreasonable at first, challenging initial conditions such as ones close to violating constraints can easily increase the problem solve time: the highest point obtained for the $N=1910$ scenario was $429$ seconds. Ideally, and specially when occupants experience thermal discomfort, the control action has to be computed in negligible time to be applied as soon as possible to the system. Keeping the solve times below a low percentile of the total sampling period is thus a common desideratum. In our application, we regarded $N=794$ to be an adequate choice.

\bibliographystyle{elsarticle-num}
\bibliography{refs.bib}

\end{document}